\documentclass[manuscript,authorversion,nonacm]{acmart}

\usepackage{comment}
\usepackage{multirow}
\usepackage{graphicx}
\usepackage{caption}
\usepackage{subcaption}
\usepackage{url}
\usepackage{xurl}
\usepackage{hyperref}
\usepackage{listings}
\usepackage{float}
\usepackage{makecell}
\usepackage{array}
\usepackage{placeins}
\usepackage{dblfloatfix}
\newcommand{\heart}{\ensuremath\varheartsuit}

\AtBeginDocument{%
  }

\begin{document}

\title{Why People Share Social Media Screenshots}

\author{Tarannum Zaki}
\affiliation{%
  \institution{Old Dominion University}
  \city{Norfolk}
  \state{Virginia} 
  \country{USA}
}
\email{tzaki001@odu.edu}

\author{Michael L. Nelson}
\affiliation{%
  \institution{Old Dominion University}
  \city{Norfolk}
  \state{Virginia} 
  \country{USA}
}
\email{mln@cs.odu.edu}

\author{Michele C. Weigle}
\affiliation{%
  \institution{Old Dominion University}
  \city{Norfolk}
  \state{Virginia}
  \country{USA}
}
\email{mweigle@cs.odu.edu}

\renewcommand{\shortauthors}{Tarannum, Nelson, Weigle}

\begin{abstract}
Posting screenshots of social media posts is a common way of increasing platform functionality, enabling inter-platform interoperability, and gives users more control over how content is presented and interpreted. While these are legitimate purposes, screenshot sharing is also used in ways that can remove context, hide the original source, or mislead audiences. We examine the motivations behind sharing screenshots of social media posts on social media platforms like Twitter/X, Instagram, and Facebook. By focusing on user motivations, our study contributes to understanding how screenshot practices shape the sharing and interpretation of information in online spaces.
\end{abstract}

\keywords{Screenshot, social media, Twitter, X, Instagram, Facebook}

\maketitle

\section{Introduction}\label{intro}
Posting screenshots of social media posts is a common way that users share content across social media platforms. Rather than relying solely on built-in sharing features, users often capture images of posts and redistribute them within or across platforms. Based on observation, we categorize screenshot use into a number of distinct but potentially overlapping functions.  First, it is a method for cross-platform interoperability (e.g., sharing a Twitter post on Facebook). Users often want to share content between platforms, but inter-platform functionality is limited.  The second general case is the sharing person anticipating the post, content or metadata, might be deleted or edited by the original poster.  The screenshot is a fixed format under the control of the sharing poster, not the original poster. A third general case is aggregation of posts.  Most social media platforms allow ``quote'' or ``reply'' functionality, but these operations are typically on a single post.  There are often cases where the sharing person wishes, in a single post of theirs, to operate on more than one post from others.  A related, fourth function is annotation: adding text, images, or commentary to the post in a manner that the original platform does not enable. 

\begin{figure*}
    \includegraphics[scale=0.4]{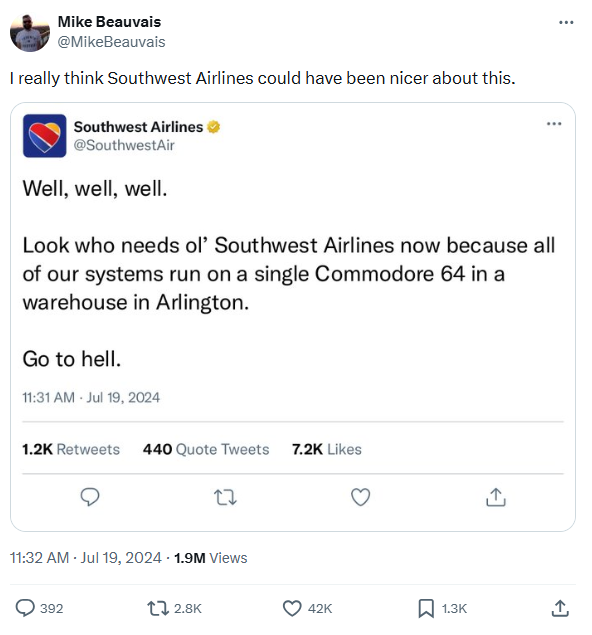}
    \Description{...}
    \caption{A fake satirical tweet that appeared to be posted by @SouthwestAir.}
    \label{satire5}
\end{figure*}

The use cases described above assume that the post(s) being shared via a screenshot actually exist. Not necessarily ``true'' or ``real'', as in they stated a true fact, but as in they existed in the social media platform with the datetime and attribution that the screenshot conveys. Sometimes people knowingly share screenshots of ``fake'' posts for humor or satire, oftentimes depending on a cultural context that helps the reader identify that the alleged poster would not have said what is attributed to them, references a scenario known to not be true, etc. Figure \ref{satire5} shows a screenshot \href{https://x.com/MikeBeauvais/status/1814322465591464077}{shared on Twitter} where a fake satirical tweet appeared to be posted by \href{https://x.com/SouthwestAir}{@SouthwestAir} went viral due to Microsoft's global outage in July 2024 \cite{v2024}. Readers familiar with the outage and discussions of Southwest's reliance on outdated computer systems could recognize the post as satire, while others might mistakenly interpret it as a genuine post from the airline. Fabricated screenshots may also be created or circulated with the intention of misleading an audience, manipulating its response, or causing harm. The same screenshot may later be shared by people who do not realize that it is fabricated. 

The boundaries between humor, unintentional false attribution, and deliberate attempt to mislead can be fuzzy and context dependent. In our research group, the polarizing issue of pineapple on pizza plagues our lunchtime gatherings. Sharing a fake post that claims ``Prof. Nelson \heart pineapple'' is humor or satire if you know him and his anti-pineapple proclivities. Someone outside the group might share the screenshot because they mistakenly believe it is real. On the other hand, a person who knows it is fake but shares it to convince the person placing the lunch order to add pineapple to every pizza would be deliberately using the screenshot to create a misleading impression and unfortunately, the pineapple juice goes everywhere.

\begin{figure*}
    \includegraphics[scale=0.4]{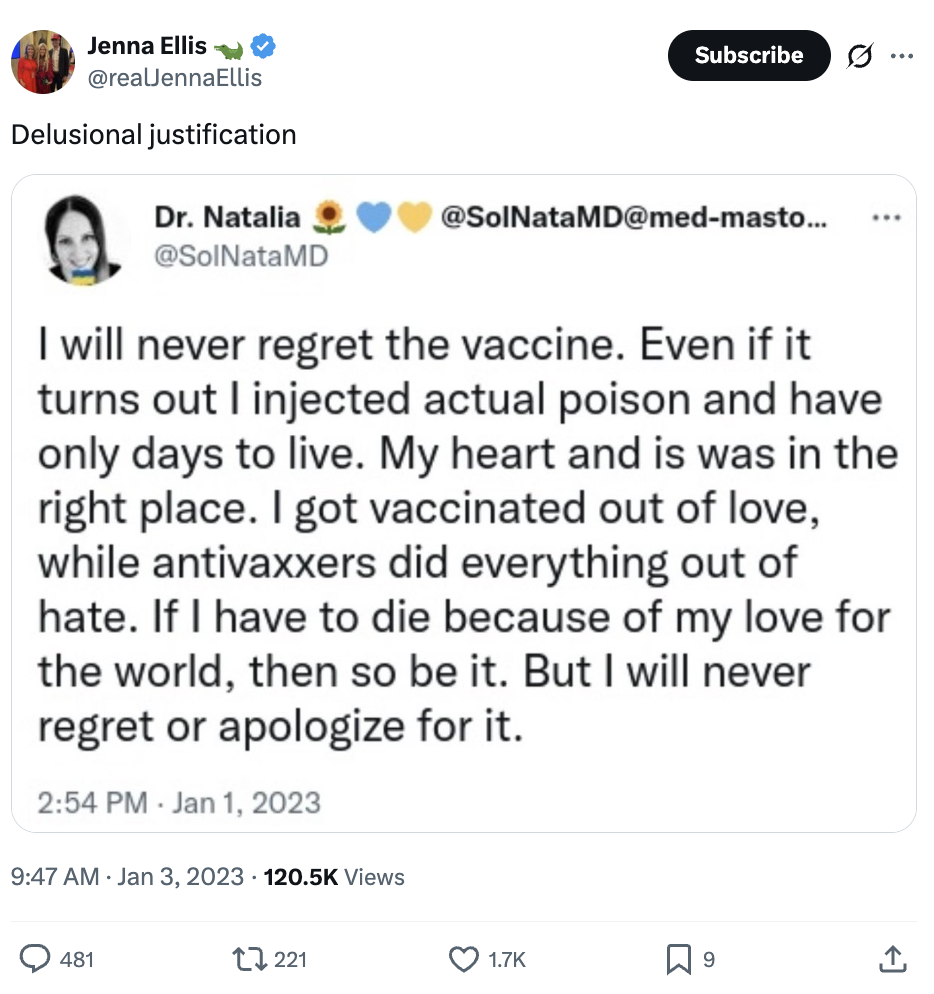}
    \Description{...}
    \caption{Screenshot of a tweet \href{https://twitter.com/jennaellisesq/status/1610286656950525952}{shared} on Twitter, falsely attributed to @SolNataMD.}
    \label{covid_tweetexample}
\end{figure*}

Sharing manipulated screenshots, knowingly or unknowingly, introduces challenges related to context loss, misinterpretation, and content authenticity. Such cases can have disastrous consequences, especially when applied to issues like public health and politics. Figure \ref{covid_tweetexample} shows an example where a screenshot of a tweet falsely attributed to a Florida doctor \href{https://x.com/SolNataMD} {Natalia Solenkova} circulated on social media. The tweet claimed that she would not regret receiving the COVID-19 vaccine even if it were fatal, leading to significant online harassment. The screenshot was later revealed to have been fabricated \cite{Trela2023}. 

Therefore, it is important to understand why people share screenshots of social media posts in order to examine social media behavior, online discourse, and the spread of information. Our goal is to categorize and discuss common reasons why social media screenshots are shared. Social media posts by public figures \cite{Harris2024}, journalists, and institutions are frequently treated as authoritative statements, making their screenshots powerful artifacts when used as evidence in news reporting, political debate, and online accountability practices. Accordingly, herein we use examples attributed to public figures to discuss the user motivations behind sharing social media screenshots.

\section{Background and Related Work}
Social media platforms rely heavily on engagement-based algorithms to determine which content is promoted or prioritized \cite{Boston2025}. Engagement metrics such as likes, comments, shares, retweets, and replies act as signals that indicate user interest, causing posts to gain increased visibility through recommendation systems and ranking feeds. Research shows that these engagement metrics not only shape what audiences see but also reward content creators by amplifying their reach \cite{Milli2025}. However, when users share screenshots of social media posts instead of using built-in sharing features, this engagement loop is effectively bypassed because the original content does not receive algorithmic boosts or notifications. Connolly \cite{Connolly2025} investigated how influential screenshots are in spreading viral content across Twitter/X, TikTok, and Instagram. He found that content becomes viral across platforms when users keep sharing it, adapting it, and giving it new meanings, rather than simply reposting the original version. As a result, screenshots enable content to spread beyond algorithmic control, disrupting the platform's ability to accurately track how information is consumed and amplified.

Screenshot sharing contributes to brigading by converting a social media post into portable visual evidence that can be circulated outside its original platform and audience. This exposure can intensify context collapse, as new audiences encounter the post without its original conversational setting. Corry \cite{Corry2021} discussed how screenshots play a crucial role in shaping moral judgment online, particularly in practices of public shaming on social media. Bigman et al. \cite{Bigman2023} examined how U.S. college students circulated race-related content on social media from relatively private spaces into larger public audiences. This cross-platform sharing of screenshots increased the visibility of racial discourse, removed it from its original audience context, and encouraged collective discussion. Lawson \cite{Lawson2021} studied how influencers, brands, and audiences managed racism-related callout campaigns in the online beauty community across YouTube, Reddit, and Twitter. He found that audiences curated screenshots and other digital ``receipts'' to preserve evidence and collectively criticized influencers and companies. In this way, screenshots can support digital vigilantism and public shaming by making a person or post highly visible, searchable, and available for collective criticism, reporting, or harassment.

Screenshots capture only the visible surface of a digital interface. This removes temporal, interactive, and contextual elements that are essential for fully understanding the captured content. Cramer et al. \cite{Cramer2023} described the static interface state of a screenshot as a ``digital window.'' They argued that screenshots should not be understood as transparent records of digital reality because the ``digital window'' shows the rendered view of content while concealing the underlying real-time process. Thus, screenshots enable de-contextualization and manipulation. Inwood and Zappavigna \cite{Inwood2024} examined how screenshots are legitimized as visual evidence in social media environments. They showed that screenshots are frequently presented in YouTube videos as proof to support misleading narratives, even though they are often taken out of context and selectively framed. 

Screenshotting is a lightweight and versatile practice enabled by operating system shortcuts or external tools for capturing, sharing, and reusing information across contexts. Sharing screenshots serves as an intentional and widespread social practice rather than an accidental byproduct of interaction. Mottelson \cite{Mottelson2023} conducted a user study to investigate what motivates people to take screenshots on their smartphones. He found that less than 40\% of screenshots originate from social media applications, highlighting that  motivations for taking screenshots extend well beyond social use. Cramer et al. \cite{Cramer2019} explored how and why people use screenshots based on the Uses and Gratifications Theory (UGT). They performed a user survey and found that the primary motivations for using screenshots are documenting (e.g., saving recipes, travel info), sharing funny social media posts, keeping evidence, and promoting activities. 

Prior research has identified several common motivations for sharing different types of screenshots on social media through specific case studies and user studies. Unlike screenshots of random information (e.g., recipes, quotes), screenshots of social media posts can shape collective understanding and influence public opinion even when the original content has been deleted, altered, or taken out of context. Our work extends existing literature towards a more technical, example-centered understanding of particularly screenshots of social media posts shared on social media.

\section{What Motivates People to Share Screenshots of Social Media Posts?}
Some common reasons that motivate social media users to share screenshots on different social media platforms are cross-platform sharing, evidence for deleted posts, aggregating screenshots, satire/humor purpose, enabling commentary and annotation, and denying engagement. We discuss these motivations using screenshot examples of Twitter/X, Facebook, and Instagram posts. We provide screenshot examples drawn from controversial yet non-political and widely recognized topics, such as the debate over ``pineapple on pizza'', posts related to the British celebrity chef Gordon Ramsay, and popular American singer Taylor Swift. These topics are useful for discussing social media screenshots because they are familiar and frequently spark engagement across platforms.

\subsection{Cross-platform sharing}
\begin{figure*}
    \includegraphics[scale=0.3]{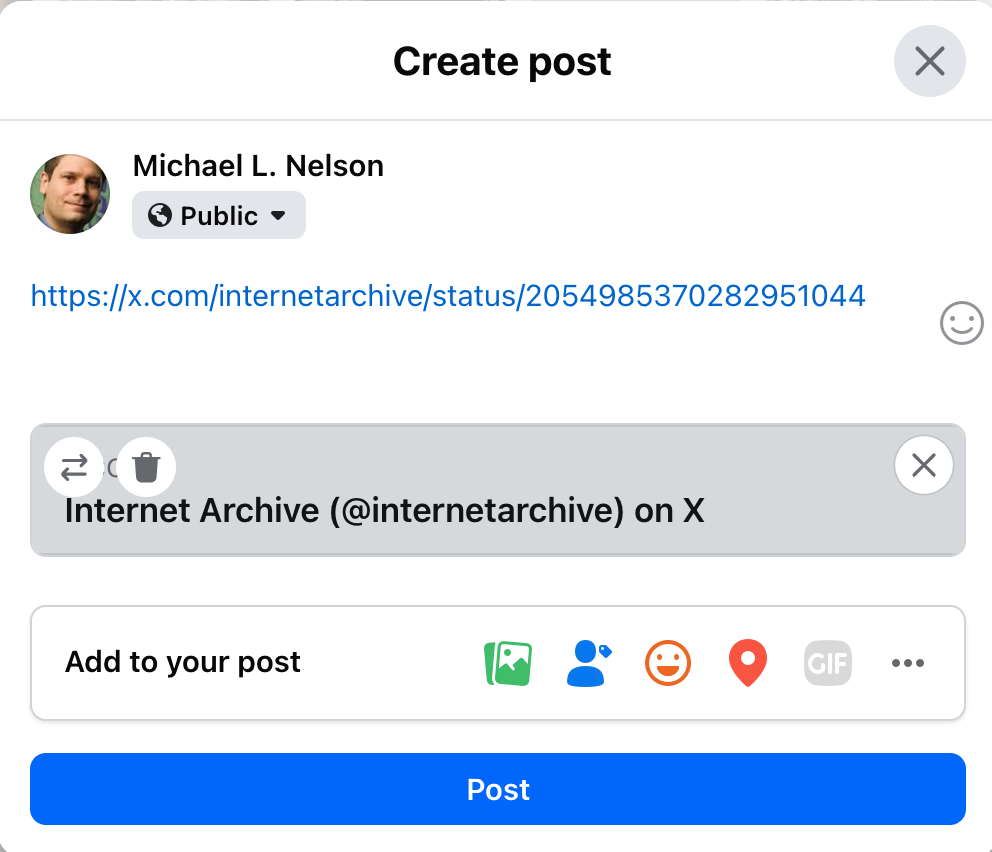}
    \Description{...}
    \caption{The purposely limited UI of inter-platform sharing (e.g., a Tweet in Facebook).}
    \label{inter_platform_sharing}
\end{figure*}

\begin{figure*}
    \includegraphics[scale=0.2]{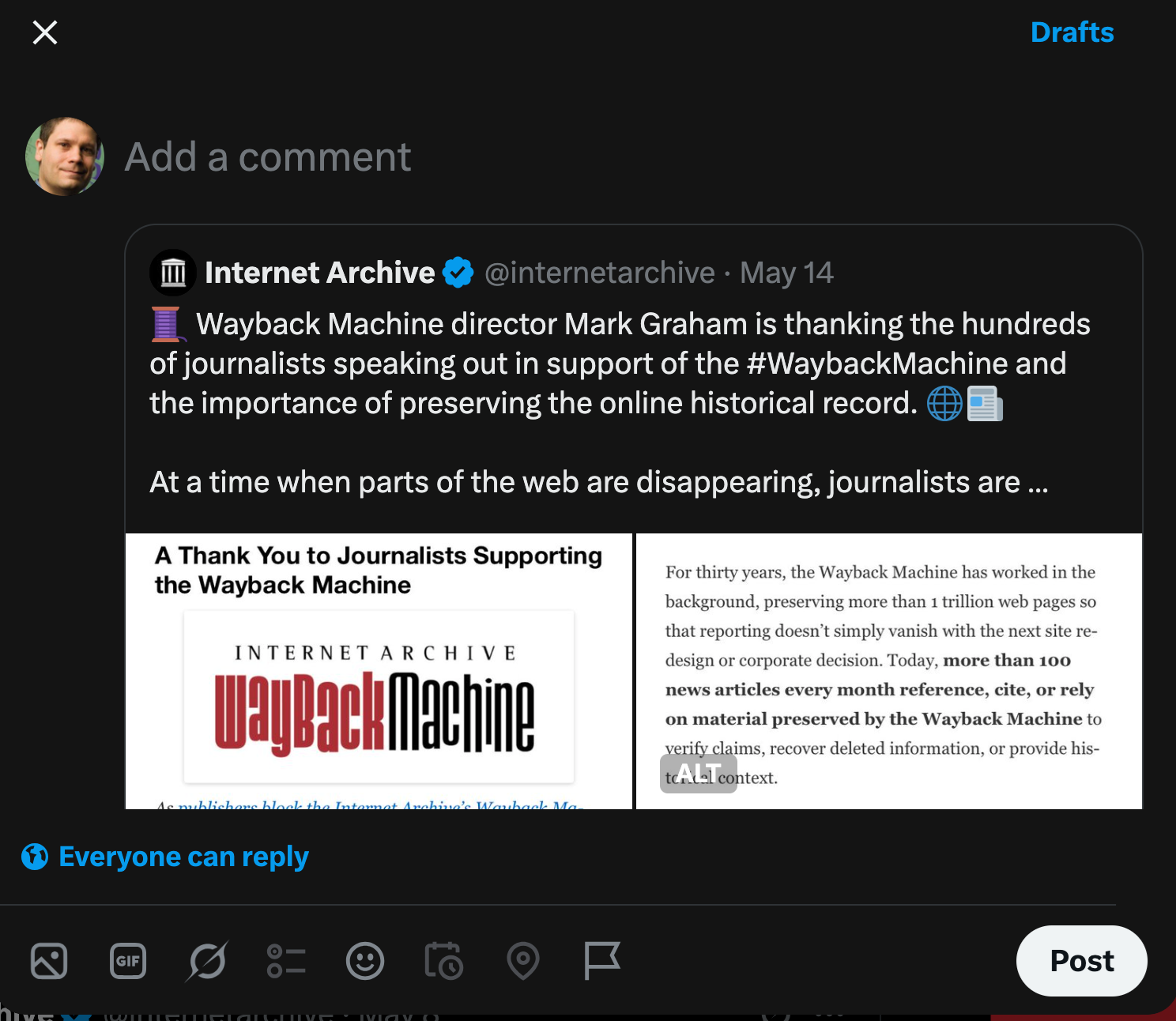}
    \Description{...}
    \caption{The richer UI of intra-platform sharing (e.g., a Tweet in Twitter). }
    \label{intra_platform_sharing}
\end{figure*}

Each social media platform has its own built-in features to allow users to interact with a post. For example, Facebook has like, react, comment, and share as features to interact with a post. Similarly, Twitter/X has reply, retweet/repost, like, bookmark, and share. It is possible to interact with a post within a specific social media platform using different internal mechanisms, but those do not allow interacting with a post on other social media platforms. For example, we can share a link of a Facebook post on Twitter, but that would not display the post, and vice-versa. For example, Figure \ref{inter_platform_sharing} shows what happens when we try to share a tweet\footnote{\url{https://x.com/internetarchive/status/2054985370282951044}} in Facebook.  The base URL is shown, and the preview is limited to the bare minimum.  Figure \ref{intra_platform_sharing} shows how the same tweet looks when shared in Twitter; the URL is replaced with a rich preview, showing both the text and images in the original tweet. Facebook \emph{could} generate the same preview of the tweet, but Facebook chooses not to in order to subtly discourage cross-platform sharing. Savvy social media users may respond with a screenshot of the tweet rather than the URL to circumvent the passive-aggressive inter-platform friction.

Screenshots are more visually appealing and easier to grasp than textual links. As a result, social media users use screenshots of posts to engage in cross-platform sharing. 
Table \ref{cross_platform_matrix} provides examples of screenshots used for cross-platform sharing between Twitter, Facebook, and Instagram.

\begin{table*}[h]
    \centering
    \caption{Summary of the images of cross-platform sharing for Twitter, Facebook, and Instagram.}
    \label{cross_platform_matrix}
    \begin{tabular}{|>{\raggedright\arraybackslash}p{3cm}|l|l|l|}
        \hline
        \multirow{2}{*}{\makecell[l]{\textbf{Platform of the} \\ \textbf{Screenshot Content}}}
        & \multicolumn{3}{c|}{\textbf{Platform where the Screenshot is Shared}} \\ \cline{2-4}
         & \textbf{Twitter} & \textbf{Facebook} & \textbf{Instagram} \\ \hline
        \textbf{Twitter} & - & Appendix \ref{app:tweetONfb} & Appendix\ref{app:tweetONinsta} \\ \hline
        \textbf{Facebook} & Appendix \ref{app:fbONtweet} & - & Appendix \ref{app:fbONinsta} \\ \hline
        \textbf{Instagram} & Appendix \ref{app:instaONtweet} & Appendix \ref{app:instaONfb} & - \\ \hline
    \end{tabular}
\end{table*}

\begin{figure*}
    \includegraphics[scale=0.45]{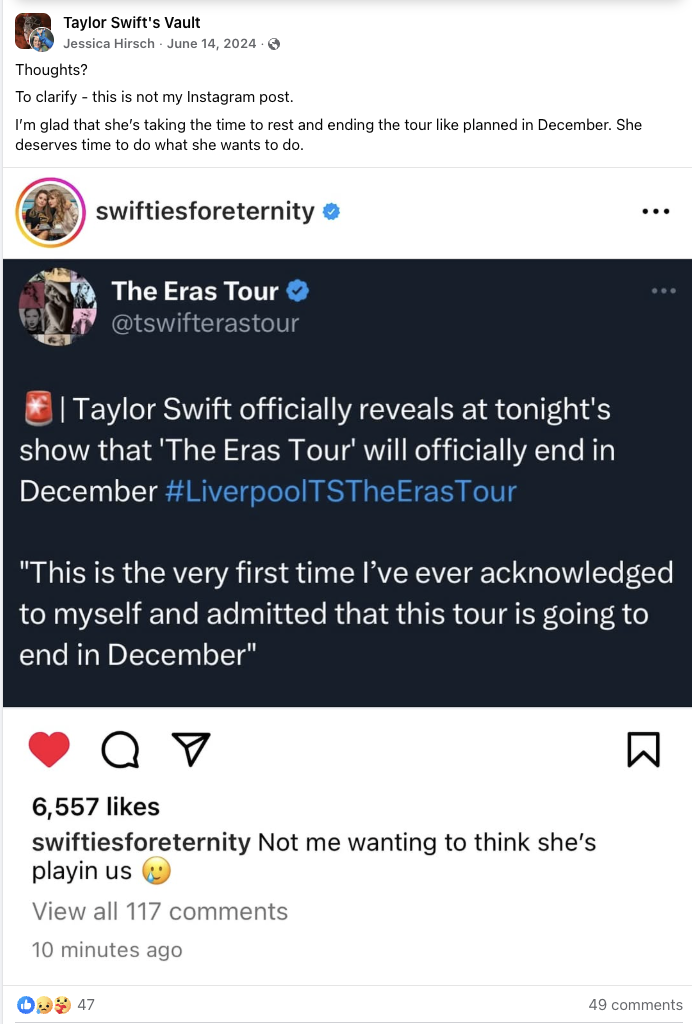}
    \Description{...}
    \caption{A Facebook account (Taylor Swift's Vault) \href{https://www.facebook.com/groups/2254218764714763/posts/3206474876155809/}{shared} a screenshot of an Instagram post which shared a screenshot of a tweet.}
    \label{nested_cross}
\end{figure*}

An even more interesting example of cross-platform sharing is the nested screenshot shown in Figure \ref{nested_cross}. The example shows a Facebook account \href{https://www.facebook.com/groups/2254218764714763/}{Taylor Swift's Vault} shared a screenshot of \href{https://www.instagram.com/swiftiesforeternity/}{swiftiesforeternity's} Instagram post. The Instagram post shared a screenshot of a Twitter account \href{https://www.x.com/tswifterastour/}{@tswifterastour's} tweet. Thus, the original Twitter content was successively embedded within Instagram and Facebook, illustrating how screenshots allow social media content to move across platform boundaries.

Users engage in cross-platform sharing to reach different groups of audiences. For example, the user base of Facebook is inclined towards an older population whereas the user base of Twitter and Instagram attracts a younger population \cite{Walton2025}. TikTok also supports cross-platform sharing by allowing users to reuse and adapt content through short videos, sounds, trends, and reactions. In the United States, Facebook is particularly common among adults ages 30 to 49, while Instagram and TikTok are especially popular among adults under 30 \cite{Pew2025}. Jayanetti \cite{Jayanetti2020} observed that American singer Katy Perry often shared the same content across Facebook, Twitter, and Instagram, suggesting that other artists may use a similar strategy to reach audiences on multiple platforms. Users' motivations also influence cross-platform sharing, as different platforms support different types of communication \cite{Petrocchi2015}. The global reach of social media platforms is another reason for cross-platform sharing. Users might want to share their post globally to boost engagement through cross-platform sharing.  Figure \ref{tweetONfb2} shows an example of how a user engages in cross-platform sharing to boost engagement. According to 2025 statistics, Facebook has a broad global reach, with particularly large user bases in regions such as Southeast Asia and Latin America \cite{ReW2025}. On the other hand, 2026 statistics show that Twitter/X has substantial user bases in countries such as the United States and the United Kingdom, with the United States accounting for the platform's largest national user base \cite{WpopRev2025}.

The social media landscape extends well beyond Facebook, Instagram, and Twitter/X, and the popularity of different platforms varies across countries and regions. Messaging services such as WhatsApp, Telegram, Signal, LINE, WeChat, and Viber also support the circulation of social media content through private messages, group chats, and broadcast channels. For example, WhatsApp is widely used across countries in South Asia, Southeast Asia, and Latin America, including India, Indonesia, Brazil, Argentina, and Mexico \cite{Pew2024}. As for regional platforms in East and Southeast Asia, WeChat is a major communication platform in China \cite{Tencent2019}, while LINE has particularly large user bases in Japan, Taiwan, and Thailand \cite{LYCorporation2026}. Screenshots may therefore move not only between public social media platforms but also through less visible messaging networks. Cinus et al. \cite{Cinus2024} demonstrated that cross-platform activity can involve coordinated sharing, identifying suspicious information-sharing patterns across Twitter/X, Facebook, and Telegram during the 2024 U.S. presidential election. Although the screenshot examples of cross-platform sharing focus on Facebook, Instagram, and Twitter/X, these platforms represent only part of a much broader global communication ecosystem.

\subsection{Evidence for deleted posts}
Users often delete posts for various professional, social, and personal reasons. Users might post something with typographical errors (``typos'') and then delete the post. Typos are not expected from public figures, and such mistakes can invite mockery or public criticism. Users might also post content that might contain unintentional errors. For example, social media accounts of celebrities, politicians, and public organizations are often handled by some representative and not by the actual account holder. The account holder may be unaware of what a representative is posting, and questionable content can be embarrassing. For example, there any number of examples of public figures or their staffers mistakenly posting or engaging with pornography via their social media accounts and then later deleting the evidence \cite{Thorbecke2016, Johnson2022, BBC2013, Taylor2017}. 

\begin{figure*}
    \includegraphics[scale=0.45]{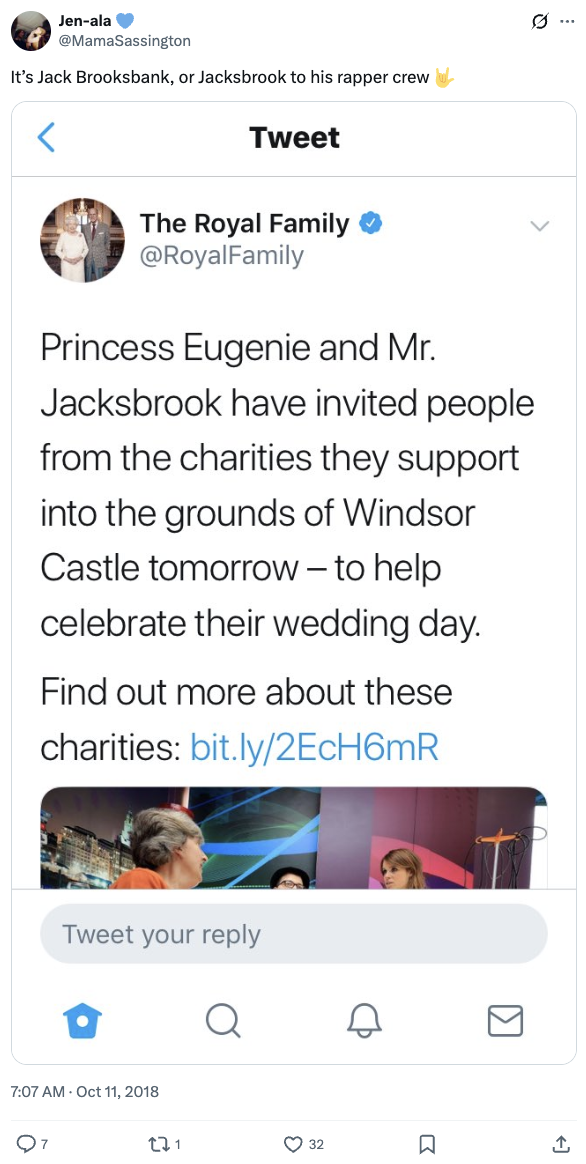}
    \Description{...}
    \caption{Screenshot of a tweet posted by @RoyalFamily having typographical and unintentional errors which was later deleted.}
    \label{delete_typo}
\end{figure*}

These are some professional reasons for deleting posts which are mostly related to career, reputation, branding, and public image. Figure \ref{delete_typo} shows an example of screenshot of a tweet \href{https://x.com/MamaSassington/status/1050341999952433154}{shared on Twitter} having typographical and unintentional errors, posted by the Twitter account \href{https://x.com/RoyalFamily}{@RoyalFamily}. The account misspelled Princess Eugenie's fiance's name Jack Brooksbank as ``Mr. Jacksbrook.'' The post was deleted and later re-posted with Brooksbank's correct name \cite{elizabeth2018}. However, the post with a typo was already shared as a screenshot on social media.

\begin{figure*}
    \includegraphics[scale=0.45]{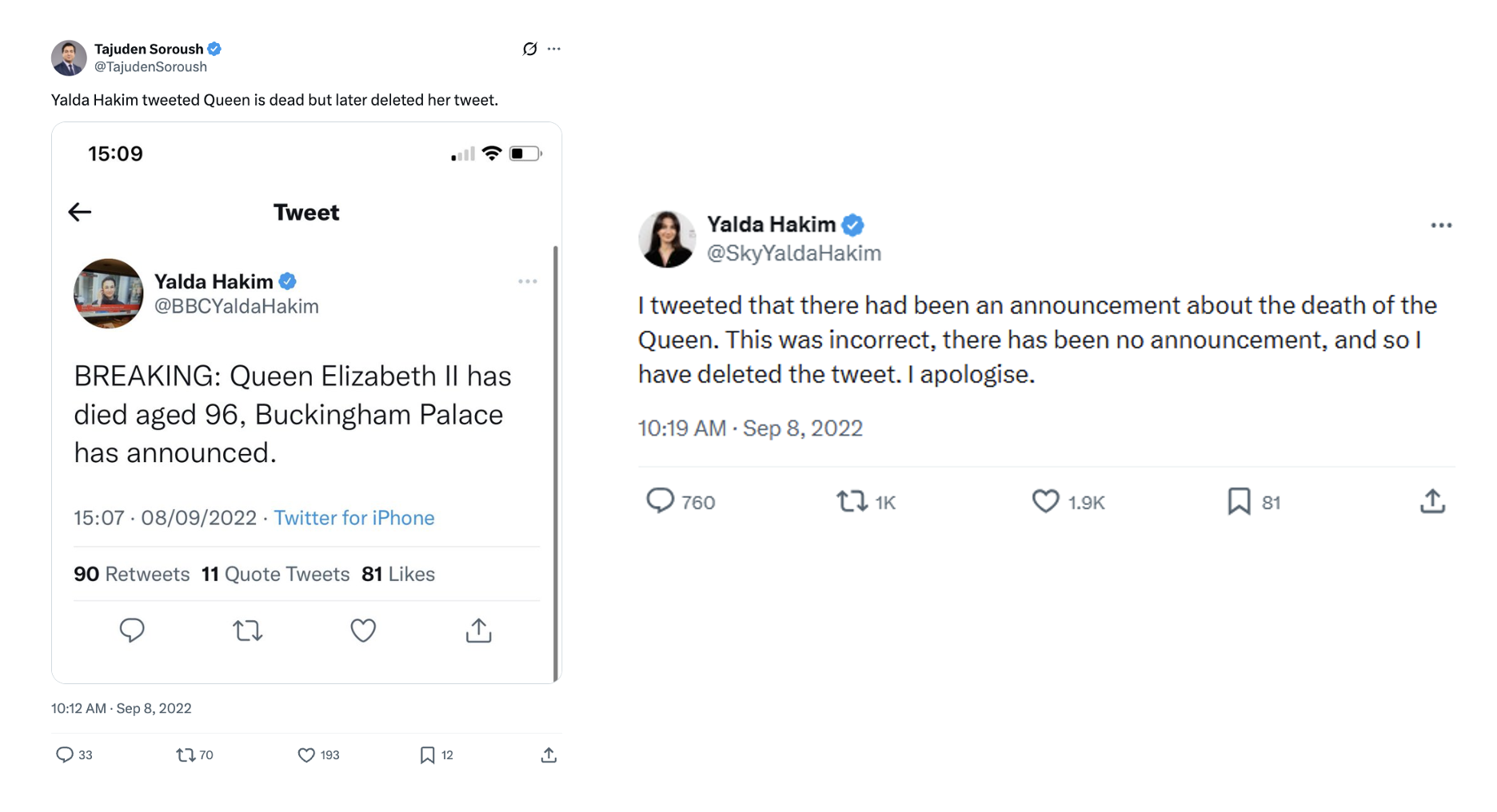}
    \Description{...}
    \caption{Screenshot of a tweet (left) posted by @SkyYaldaHakim (Yalda Hakim) having incorrect information about Queen Elizabeth's death and another tweet (right) by her showing the post was later deleted.}
    \label{incorrect_BBC}
\end{figure*}

Users might also delete posts which are socially inappropriate, emotionally triggering, poorly timed, or posted in the heat of the moment. Figure \ref{incorrect_BBC} shows screenshot of a tweet \href{https://x.com/TajudenSoroush/status/1567878481052065794}{shared on Twitter}, posted by \href{https://x.com/SkyYaldaHakim}{@SkyYaldaHakim (Yalda Hakim)} with incorrect information about Queen Elizabeth’s death. The former BBC presenter apologized and later deleted the tweet \cite{Heslop2022}, however, screenshots were already being shared.

\begin{figure*}
    \includegraphics[scale=0.45]{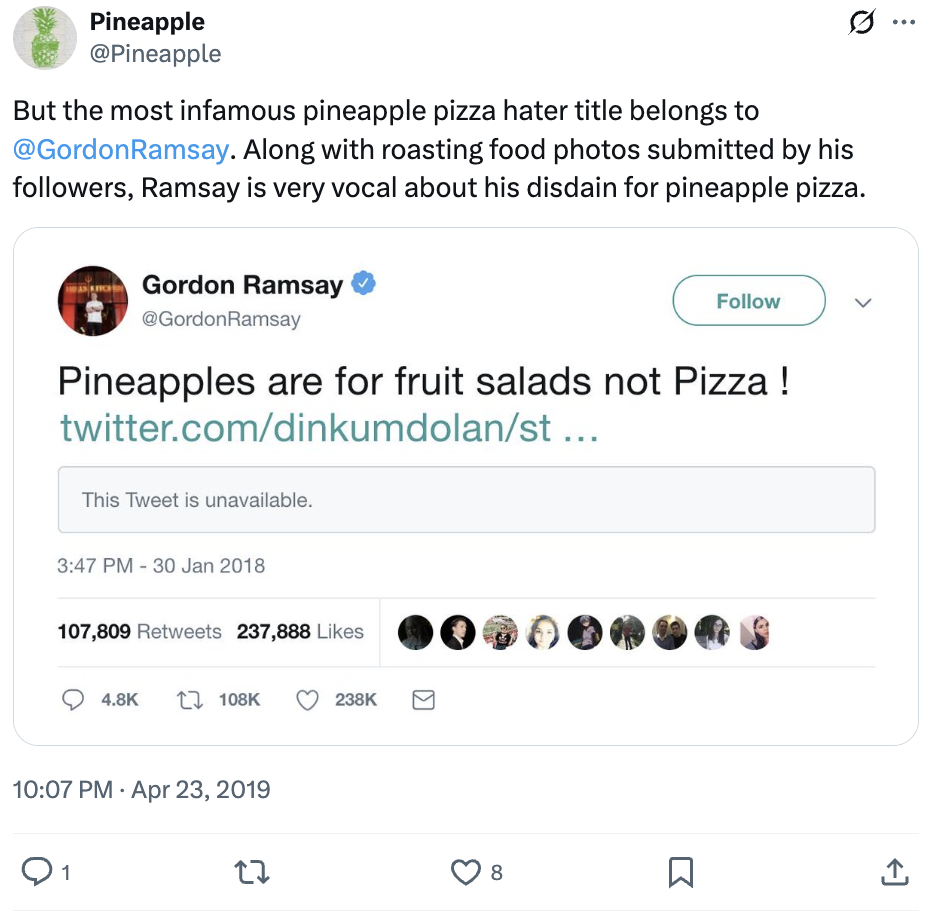}
    \Description{...}
    \caption{Screenshot of a tweet showing the embedded tweet was deleted.}
    \label{social_delete2}
\end{figure*}

Users share or post their opinions on social media platforms, but sometimes such posts might attract criticism or harassment, prompting users to delete them to stop further conflict. Ringel and Davidson \cite{Ringel2020} found that journalists deleted tweets to minimize risk and regain control in what they describe as a harassment-ridden public sphere. Minaei et al. \cite{Minaei2022} found that some social media users deleted posts specifically because they had caused controversy or harassment. Figure \ref{social_delete2} is another version of the screenshot of the tweet in Figure \ref{tweetONfb1}, which shows that the \href{https://x.com/dinkumdolan/status/958478370278387713}{embedded (quoted) tweet} was deleted. Being quote tweeted by the celebrity chef Gordon Ramsay, maybe the user thought the post was attracting unwanted criticism and decided to delete it. 

\begin{figure*}
    \includegraphics[scale=0.35]{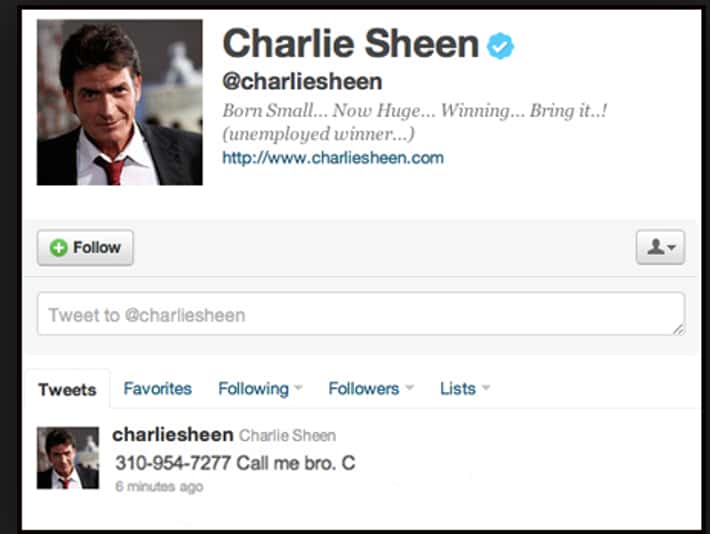}
    \Description{...}
    \caption{Screenshot of a tweet showing Charlie Sheen's personal phone number was revealed which was later deleted.}
    \label{charlie_tweet}
\end{figure*}

Users might delete posts that conflict with their personal beliefs or identity over time. Users might delete posts that might disclose more personal information than the user intended. Figure \ref{charlie_tweet} shows a screenshot of a tweet posted by \href{https://x.com/charliesheen}{Charlie Sheen (@charliesheen)} which revealed his personal phone number. The tweet was deleted almost immediately, but not before screenshots were captured and shared \cite{scoop2011}.

There are a variety of reasons why users delete their social media posts, and other users have an awareness of this risk, so they often create screenshots as a hedge against future deletion.  This transfers the role of custodianship of the post from the original author to one or of the readers.

\subsection{Aggregating screenshots}
\begin{figure*}
    \includegraphics[scale=0.45]{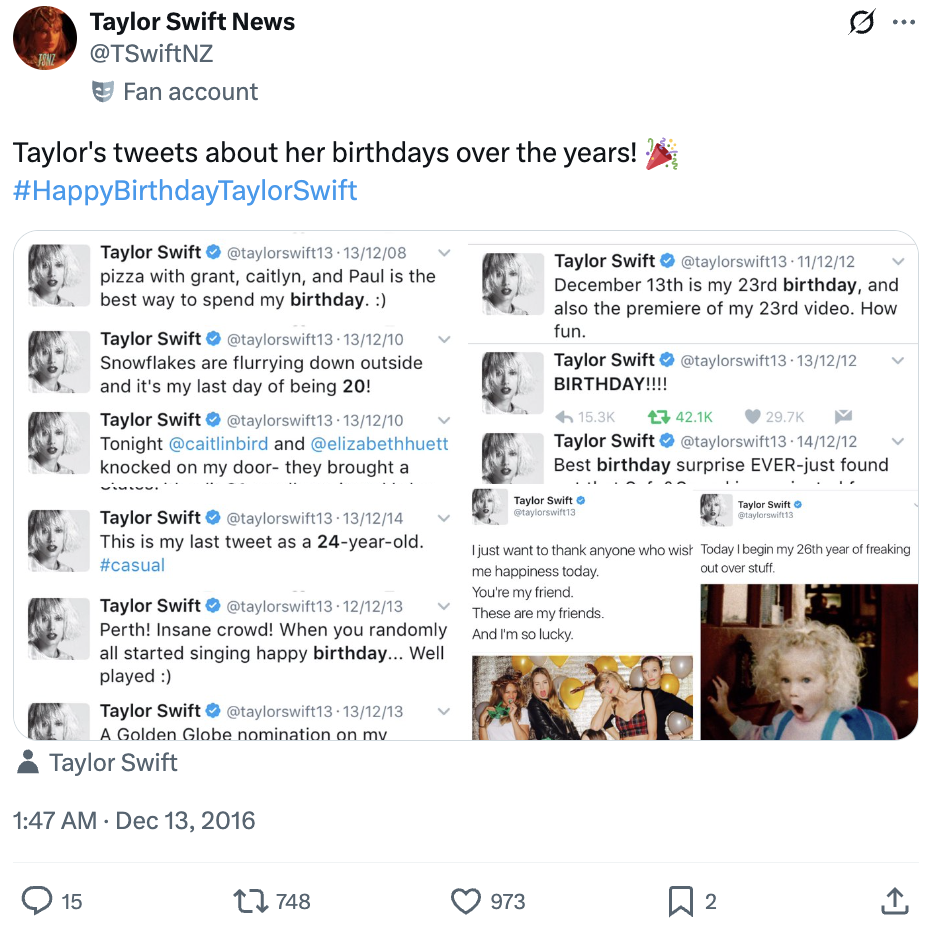}
    \Description{...}
    \caption{A screenshot showing Taylor Swift's multiple tweets stitched together and shared on Twitter.}
    \label{taylor_up}
\end{figure*}

Sometimes, people stitch together a number of screenshots based on a particular context to guide interpretation and attach to their social media posts. People do this often to present a stronger argument, to build a storyline/narrative, or to gather evidence for a certain context.  This extends the functionality of the social media platform itself, which typically only allows operations (e.g., quote, reply) on the granularity of a single post. Figure \ref{taylor_up} shows an example of screenshot \href{https://x.com/TSwiftNZ/status/808564014062018560}{shared on Twitter} consisting of Taylor Swift's multiple tweets stitched together. More examples are provided in Appendix \ref{app:aggregate_ss}.

\subsection{Satire/humor purpose}
\begin{figure*}
    \includegraphics[scale=0.3]{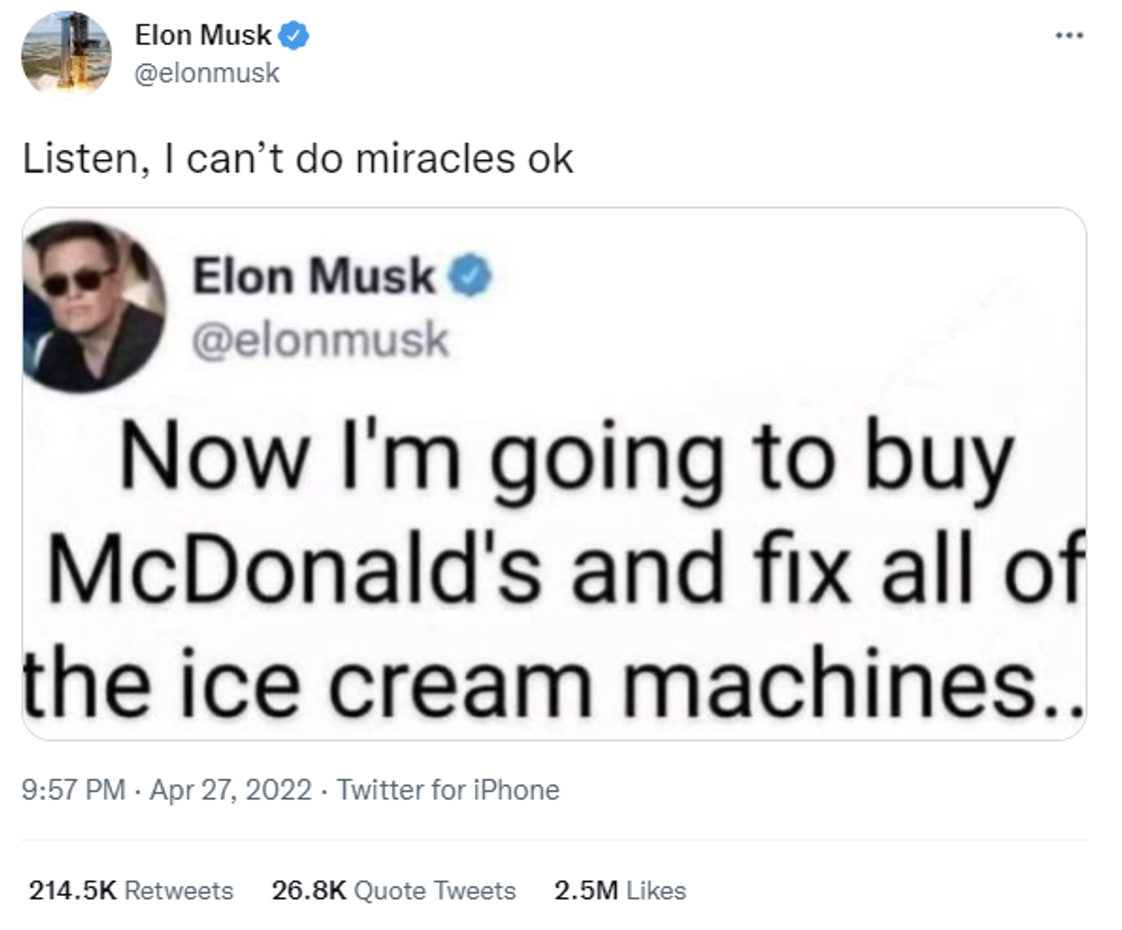}
    \Description{...}
    \caption{A fake satirical tweet appeared to be posted by Elon Musk.} 
    \label{fake-not-found}
\end{figure*}

Sometimes people construct ``fake'' social media posts for the purpose of humor or satire, with the assumption that those in the same social context will immediately recognize fabricated posts for what they are (cf. intentionally misleading as discussed in Section \ref{intro}).  Since one cannot quote, retweet, or reply to a post that never actually existed, a screenshot of the (fake) post is the only way it can exist. Figure \ref{fake-not-found} shows an example of shared screenshot of a tweet that was not posted by the alleged author. We did not find any evidence on the live web or in the web archives to support that this tweet was really posted by @elonmusk. Rather, this is one of the many satirical tweets that began to circulate when Elon Musk bought Twitter in April 2022. More examples are provided in Appendix \ref{app:satire/humor}.

\subsection{Enabling commentary and annotation}
\begin{figure*}
    \includegraphics[scale=0.45]{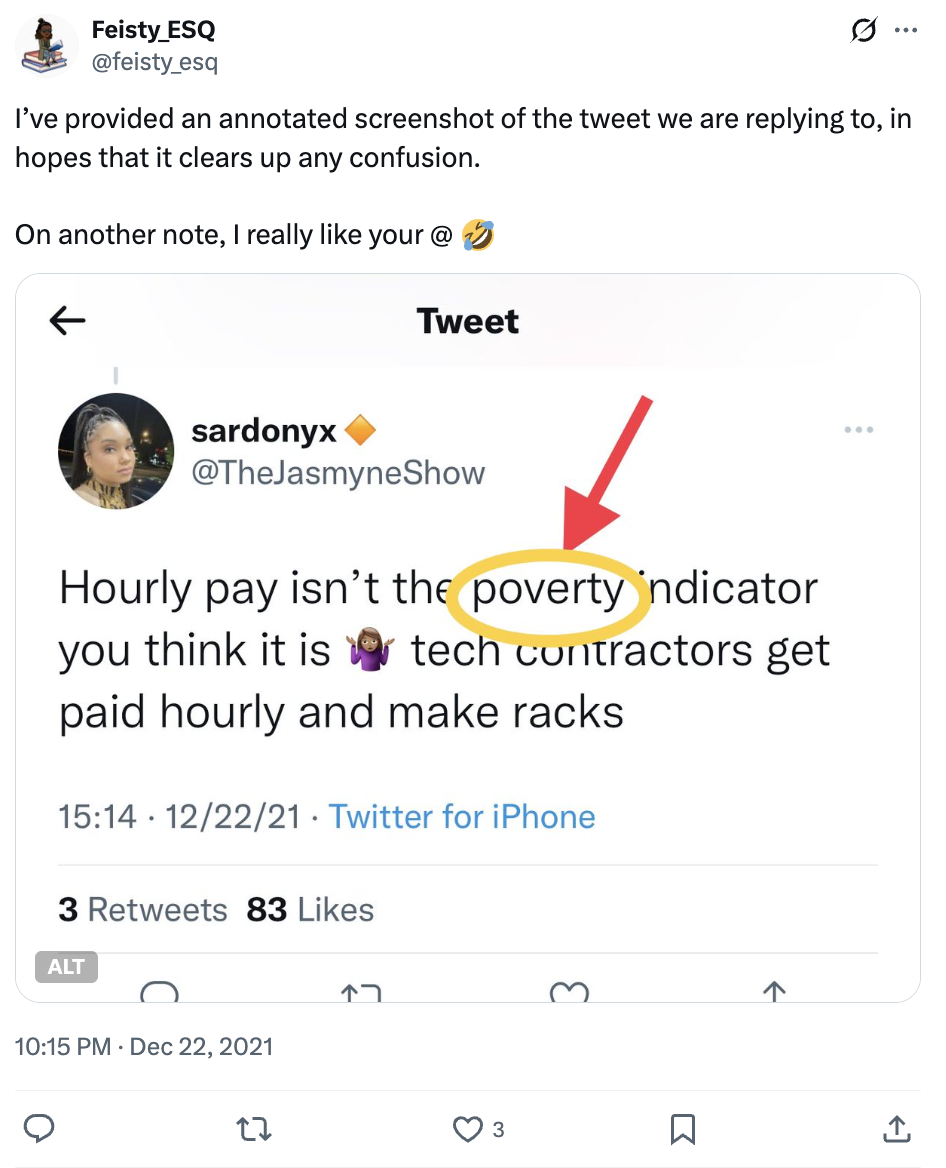}
    \Description{...}
    \caption{Screenshot of a tweet \href{https://x.com/feisty_esq/status/1473854691578728454}{shared} on Twitter, annotated with a yellow colored circle and red pointed arrow highlighting the word ``poverty'' in the tweet.}
    \label{red}
\end{figure*}

Screenshots enable commentary, watermarks, and annotations in ways not defined by the platform themselves. Users can highlight, crop, or remove specific parts of a post to draw attention to particular details or arguments. Figure \ref{red} shows screenshot of a tweet shared on Twitter, annotated with a yellow colored circle and red pointed arrow highlighting the word ``poverty'' in the tweet. More examples are provided in Appendix \ref{app:annotate}.

\subsection{Denying Engagement}
\begin{figure*}
    \includegraphics[scale=0.45]{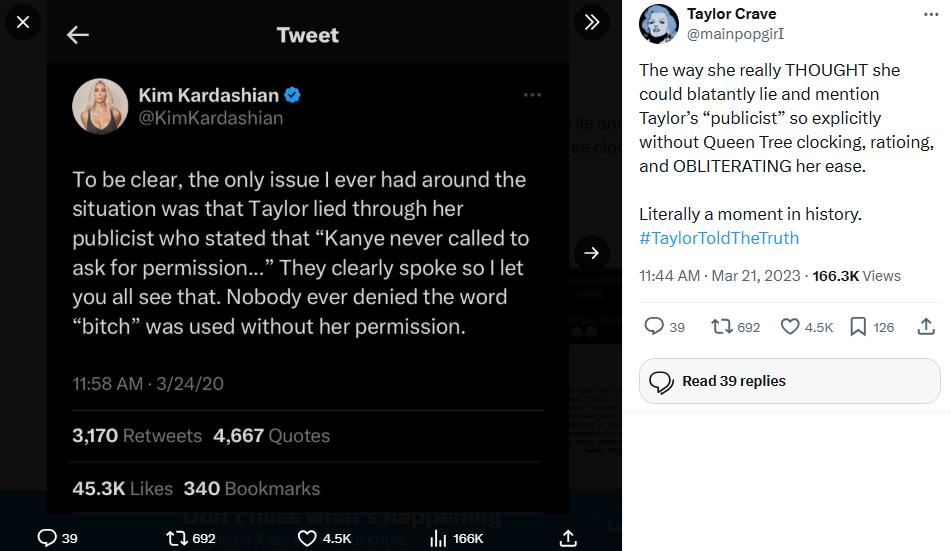}
    \Description{...}
    \caption{By \href{https://x.com/mainpopgirI/status/1638204942812057600}{sharing} a screenshot of @KimKardashian's tweet, @mainpopgirlI criticized the author's tweet without directly engaging with the original post.}
    \label{deny_kim}
\end{figure*}

\begin{figure*}
    \includegraphics[scale=0.45]{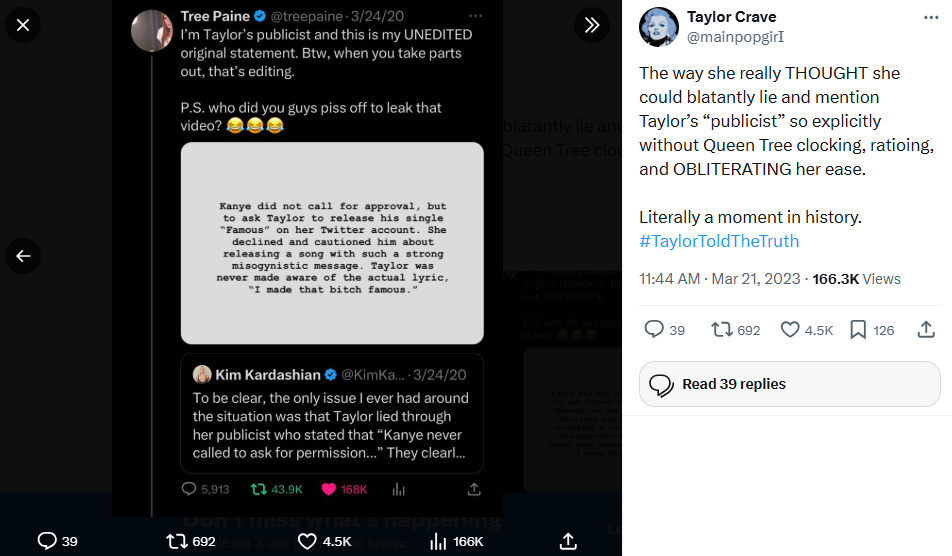}
    \Description{...}
    \caption{By \href{https://x.com/mainpopgirI/status/1638204942812057600}{sharing} a screenshot of @treepaine's tweet, @mainpopgirlI supported the author's tweet without directly engaging with the original post.}
    \label{deny_taylor_publicist}
\end{figure*}

Screenshots provide a way to engage \textit{around} content rather than \textit{with} it. This practice of indirectly sharing content enables users to withhold from the original poster engagement signals such as views, likes, shares, or algorithmic boosts. People use screenshots for denying engagement for several reasons. Users may want to criticize or discuss a post without amplifying its visibility or rewarding the original poster with additional reach. This is common when the content is controversial. Screenshots allow users to engage with the posts without notifying the original poster, reducing the risk of unwanted attention, including retaliation and brigading. Figures \ref{deny_kim} and \ref{deny_taylor_publicist} show a Twitter user \href{https://x.com/mainpopgirI}{@mainpopgirI} sharing two screenshots -- one of \href{https://x.com/KimKardishian}{Kim Kardashian (@KimKardashian)} and another of Taylor Swift's publicist \href{https://x.com/treepaine}{Tree Paine (@treepaine)}. Both authors of the screenshots are public figures. The screenshot sharer criticizes @KimKardashian's tweet while expressing support for @treepaine's tweet. Rather than quote-tweeting or replying directly, the user responds by sharing screenshots, thereby avoiding platform engagement features. In this way, the original poster (screenshot's author) is neither notified nor the original tweet (screenshot's tweet) is boosted. Thus, by using screenshots, the screenshot sharer bypasses platform-specific engagement features and deliberately avoids unwanted attention or direct confrontation with the original tweet authors. More examples are provided in Appendix \ref{app:deny_engage}.


\section{Conclusion}
While users might not explicitly think of them as such, screenshots are a way of expanding the functionality of a social media platform.  Social media platforms purposefully provide low interoperability, so users use screenshots to preserve the look and feel of the original post when posting to a different platform.  Users understand that posts, especially controversial ones, may be edited or deleted, so they screenshot the original post as a form of preserving the original form or expression.  The granularity of operations like quote and reply are on a single post, and users often want to aggregate multiple posts into one, and screenshots are the only method to do so. Fabricated screenshots are a way of providing humorous or satirical commentary, although the line between humor or satire and  unintentional false attribution, or deliberate attempt to mislead is often fuzzy. Screenshots also allow a level of non-textual commentary or annotation that the original platforms do not provide.  Finally, screenshots are an adversarial or defensive method for commentators to deny engagement to the original poster, or to comment without notification. 

By examining the motivations behind sharing screenshots, our work highlights how screenshots can significantly influence online discourse and the spread of information. Screenshots of social media posts are especially influential because of different social media platform's role in politics, journalism, and public accountability. Understanding these motivations is essential for addressing problems related to content authenticity and misinterpretation. Thus, our work provides a foundation for future research on content verification and digital literacy efforts that would help users better evaluate screenshots of social media posts in a highly visual and cross-platform media environment.

\begin{acks}
This work supported in part by the GROW M\&S project (Grant \# 300747-010), funded by the US Department of Education.
\end{acks}

\bibliographystyle{ACM-Reference-Format}
\bibliography{sample-base}

\FloatBarrier
\clearpage
\onecolumn
\appendix

\section{Screenshots of a Tweet shared on Facebook}
\label{app:tweetONfb}
\begin{figure}[H]
    \includegraphics[scale=0.4]{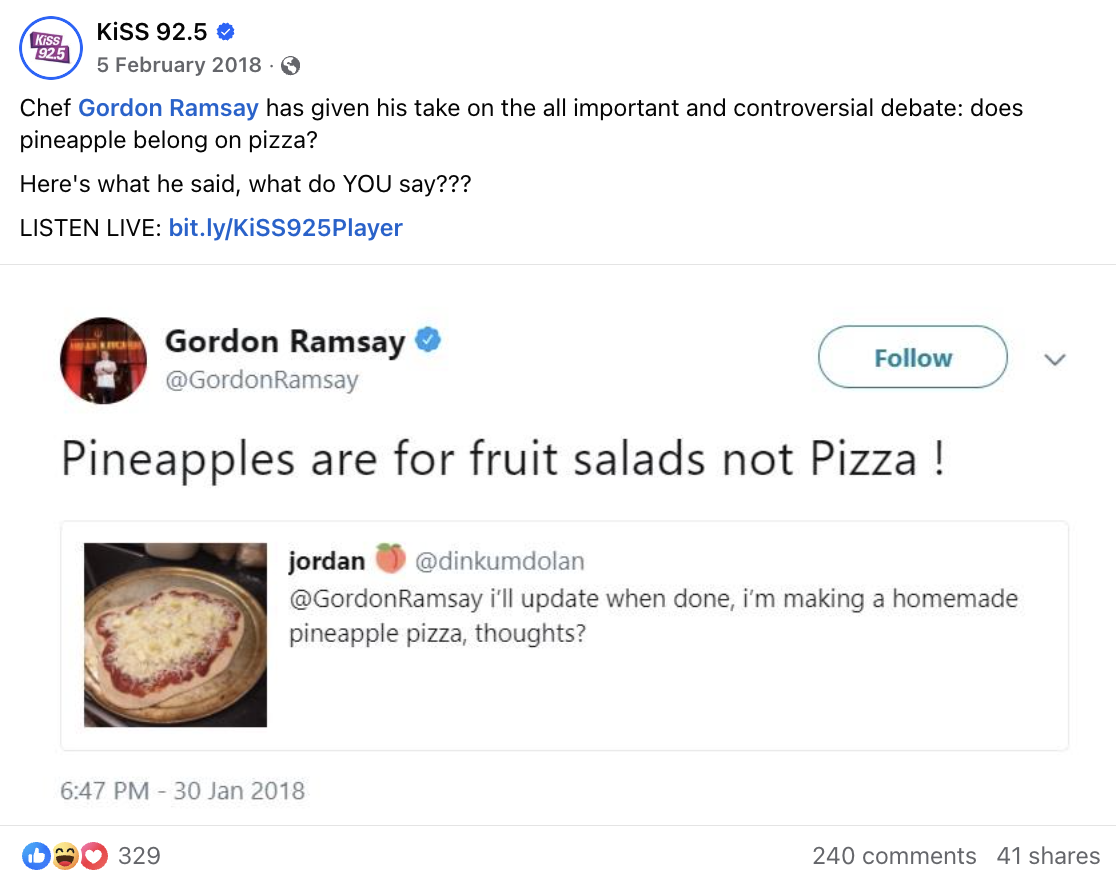}
    \Description{...}
    \caption{A Toronto radio station (KiSS 92.5) \href{https://www.facebook.com/Kiss925/posts/pfbid02cGmFiQ5vvzXT2SfLUq6BZpaZV6WQMk3TGB2y8DCtMZk423v9DfQbnjwTpKcop2eGl}{shared} a screenshot of Gordon Ramsay's tweet on their Facebook page.}
    \label{tweetONfb1}
\end{figure}

\begin{figure}[H]
    \includegraphics[scale=0.4]{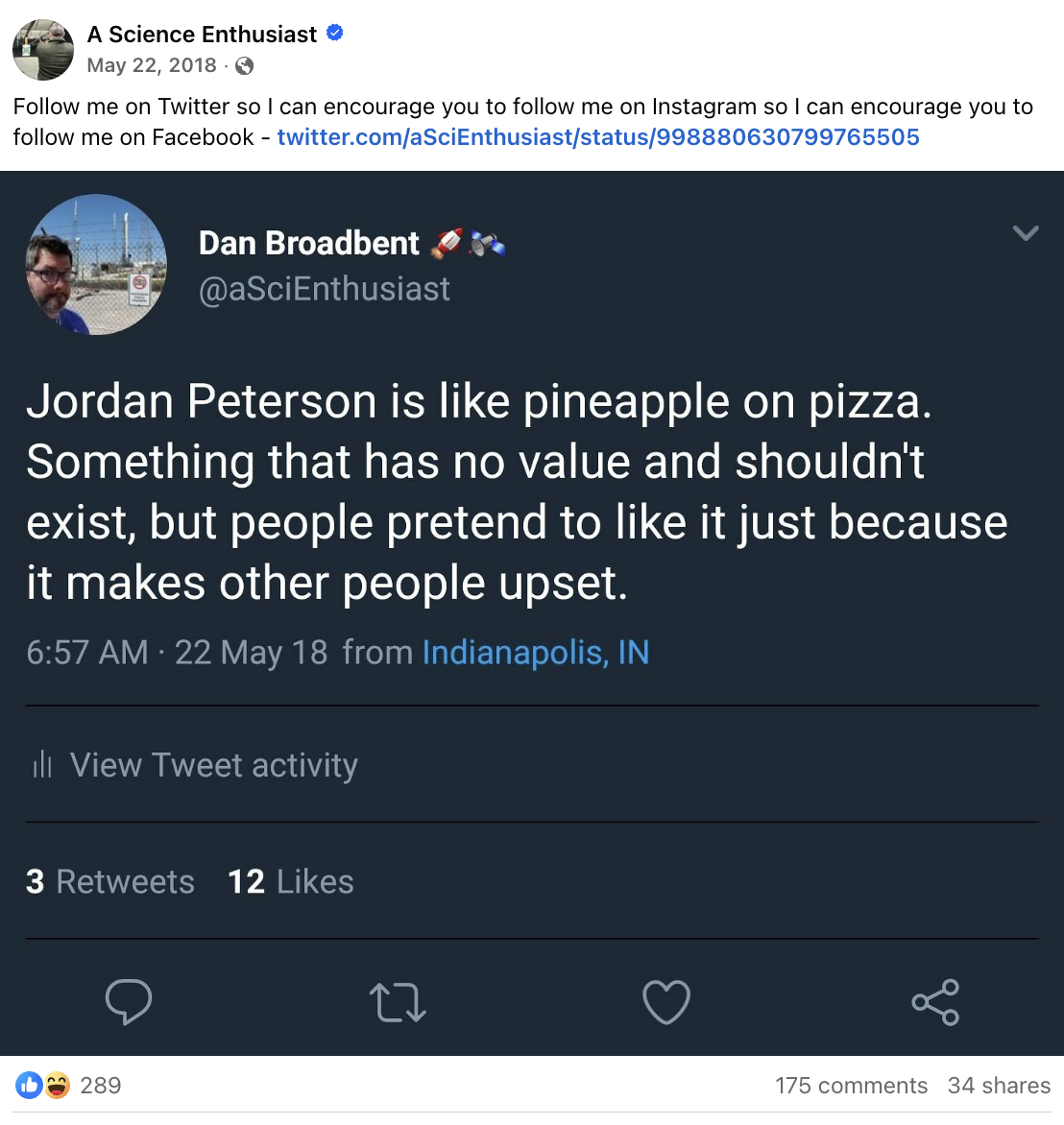}
    \Description{...}
    \caption{A Facebook account (A Science Enthusiast) \href{https://www.facebook.com/aScienceEnthusiast/posts/pfbid02DcSV7Lv2TdfWTHwAZpy9vhAaywQF7uPvkzP9N6JfPAvVQKd3DZ5xJGZfFRELzYMKl}{shared} a screenshot of his own tweet on his own Facebook page.}
    \label{tweetONfb2}
\end{figure}

\begin{figure}[H]
    \includegraphics[scale=0.4]{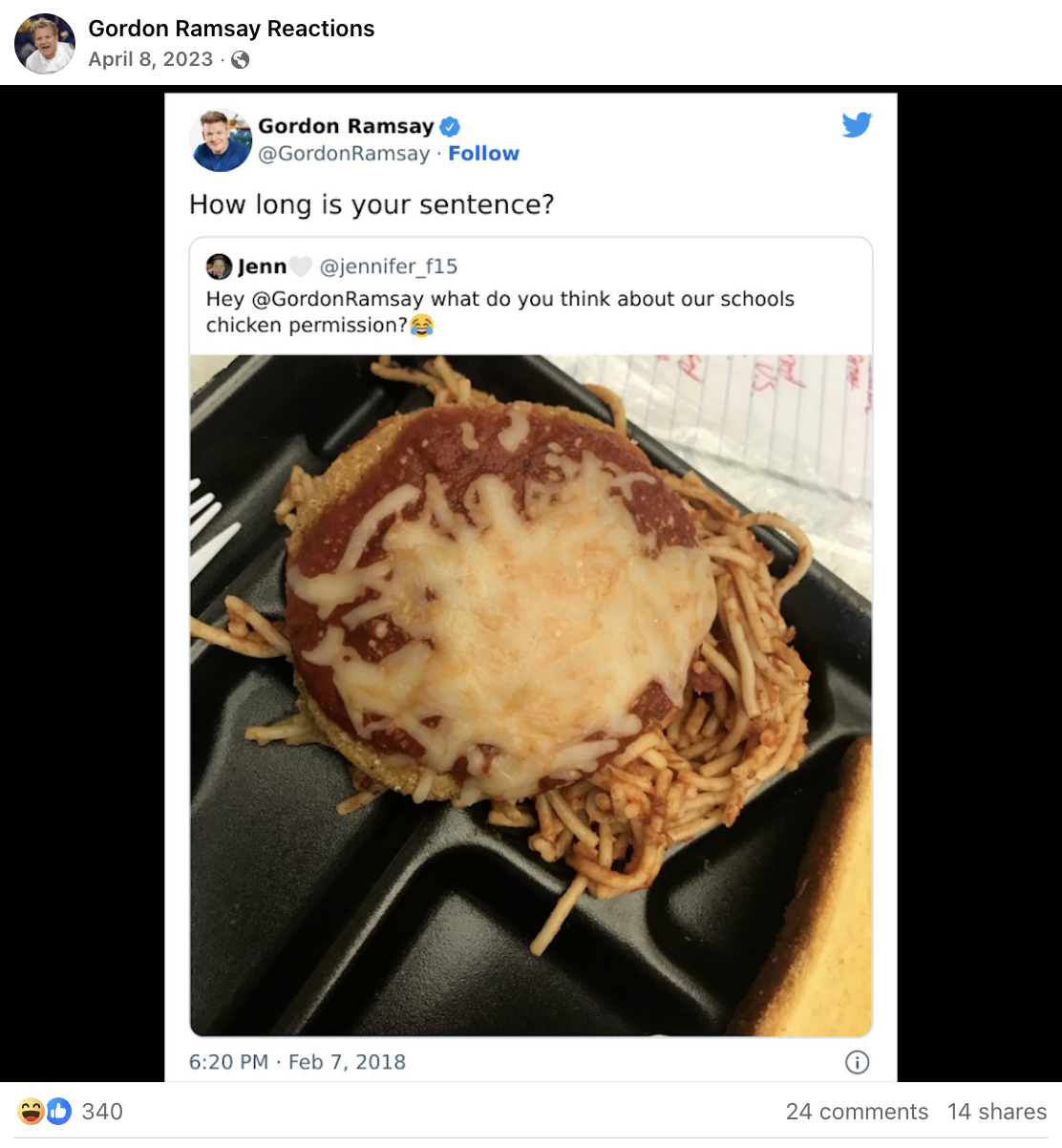}
    \Description{...}
    \caption{An Facebook account (Gordon Ramsay Reactions) \href{https://www.facebook.com/GordonRamsayReactions/posts/pfbid02oW7WDAPYBfyrBThC5R3JX9QWKnciK1RApcLt1xRexZw4xv3BTMQjKGgnmiu6Yyhvl}{shared} a screenshot of Gordon Ramsay's tweet on their Facebook page.}
    \label{tweetONfb3}
\end{figure}

Figure \ref{tweetONfb1} shows an example where a Toronto radio station \href{https://www.facebook.com/Kiss925}{(KiSS 92.5)} shared a screenshot of  \href{https://x.com/GordonRamsay}{Gordon Ramsay's} tweet on their Facebook page. Figure \ref{tweetONfb2} shows another example where a Facebook account \href{https://www.facebook.com/aScienceEnthusiast}{(A Science Enthusiast)} shared a screenshot of his own tweet on his Facebook page. Figure \ref{tweetONfb3} shows another example where a Facebook account \href{https://www.facebook.com/GordonRamsayReactions}{Gordon Ramsay Reactions} shared a screenshot of \href{https://x.com/GordonRamsay}{Gordon Ramsay's} tweet on their Facebook page.

\FloatBarrier
\clearpage

\section{Screenshots of Tweet shared on Instagram}
\label{app:tweetONinsta}
\begin{figure}[H]
    \includegraphics[scale=0.4]{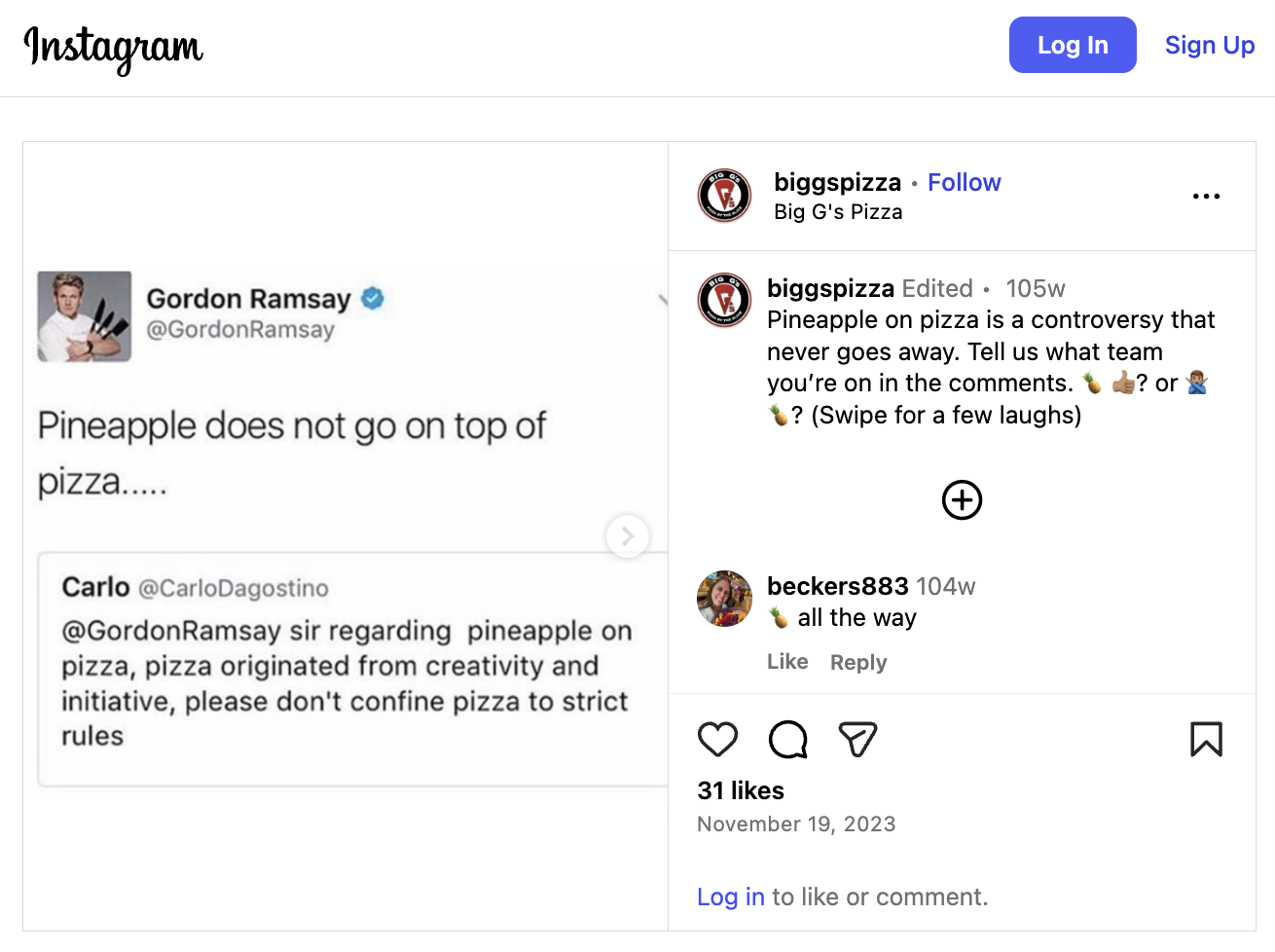}
    \Description{...}
    \caption{A Chicago pizza shop (biggspizza) \href{https://www.instagram.com/p/Cz1kVMNLqyK/}{shared} a screenshot of Gordon Ramsay's tweet on their Instagram page.}
    \label{tweetONinsta1}
\end{figure}

\begin{figure}[H]
    \includegraphics[scale=0.35]{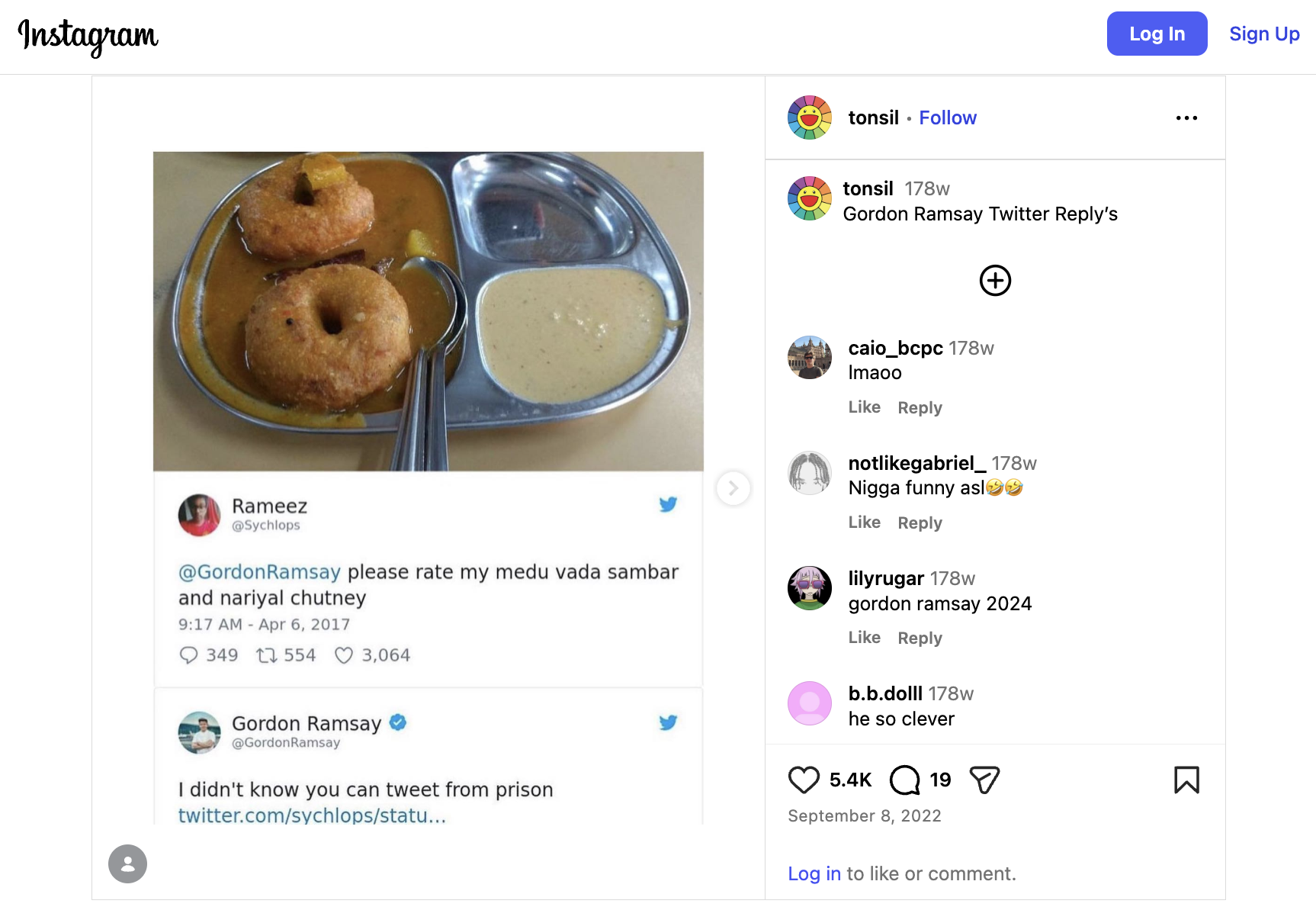}
    \Description{...}
    \caption{An Instagram account (tonsil) \href{https://www.instagram.com/p/CiQmVWNOvNq/}{shared} a screenshot of Gordon Ramsay's tweet on their Instagram page.}
    \label{tweetONinsta2}
\end{figure}

\begin{figure}[H]
    \includegraphics[scale=0.35]{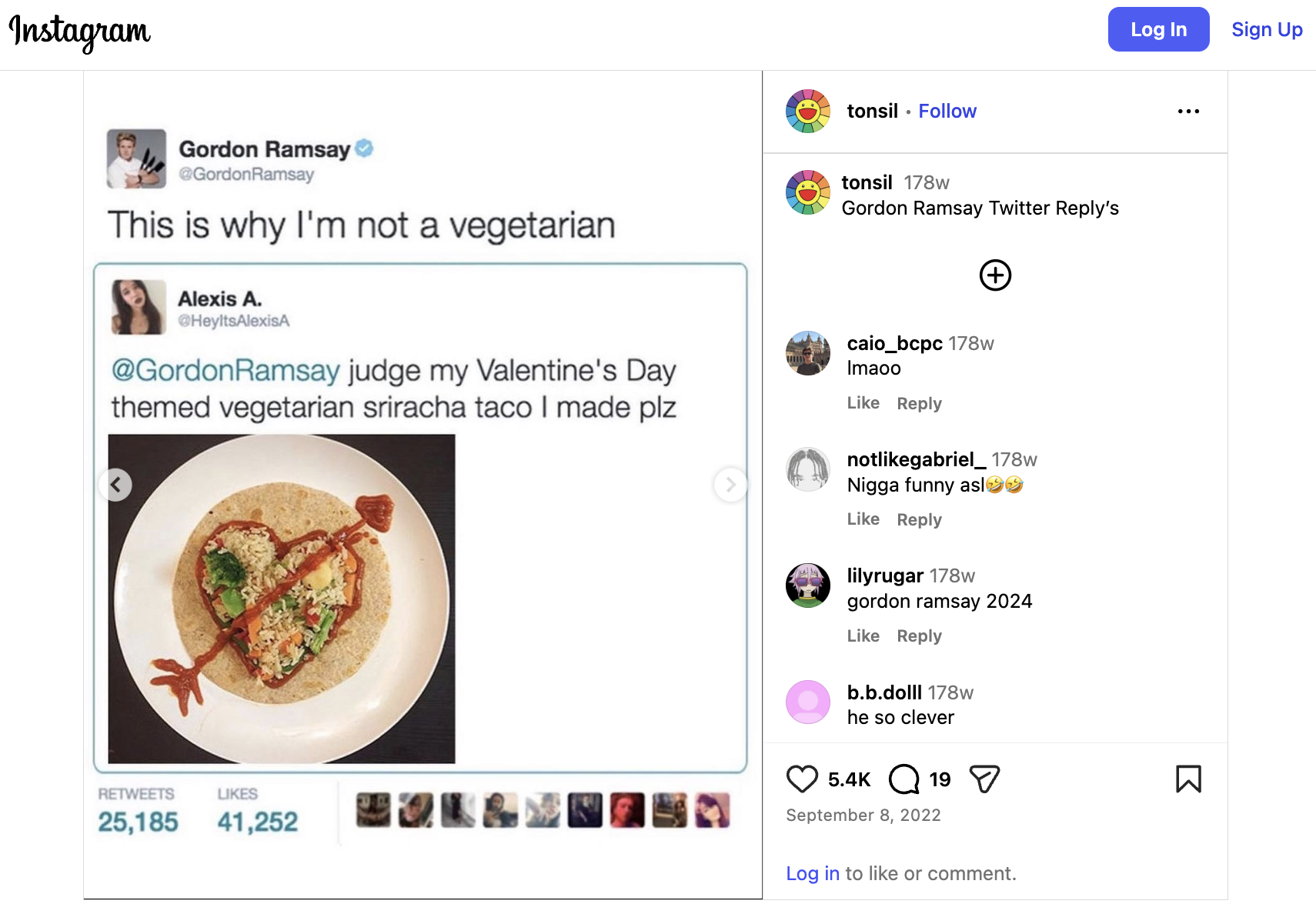}
    \Description{...}
    \caption{An Instagram account (tonsil) \href{https://www.instagram.com/p/CiQmVWNOvNq/?img_index=5}{shared} a screenshot of Gordon Ramsay's tweet on their Instagram page.}
    \label{tweetONinsta3}
\end{figure}

Figure \ref{tweetONinsta1} shows an example where a Chicago pizza shop \href{https://www.instagram.com/biggspizza}{(biggspizza)} shared a screenshot of \href{https://x.com/GordonRamsay}{Gordon Ramsay's} tweet on their Instagram page. Figures \ref{tweetONinsta2} and \ref{tweetONinsta3} show examples where an Instagram account \href{https://www.instagram.com/tonsil/?g=5}{(tonsil)} shared screenshots of \href{https://x.com/GordonRamsay}{Gordon Ramsay's} tweet.

\FloatBarrier
\clearpage

\section{Screenshots of Facebook post shared on Twitter}
\label{app:fbONtweet}
\begin{figure}[H]
    \includegraphics[scale=0.45]{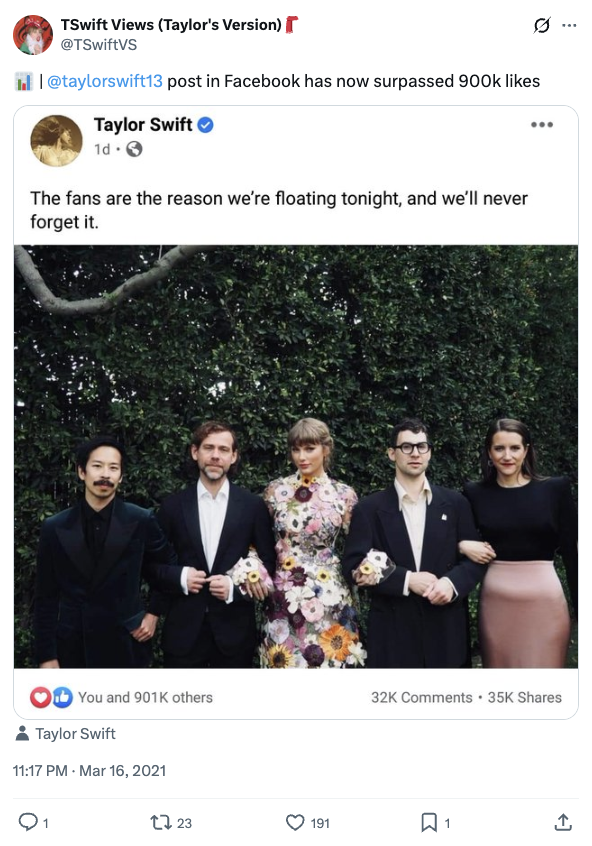}
    \Description{...}
    \caption{A Twitter account (@TSwiftVS) \href{https://x.com/TSwiftVS/status/1372024149145649154}{shared} a screenshot of Taylor Swift's Facebook post on their Twitter page.}
    \label{fbONtweet3}
\end{figure}

\begin{figure}[H]
    \includegraphics[scale=0.43]{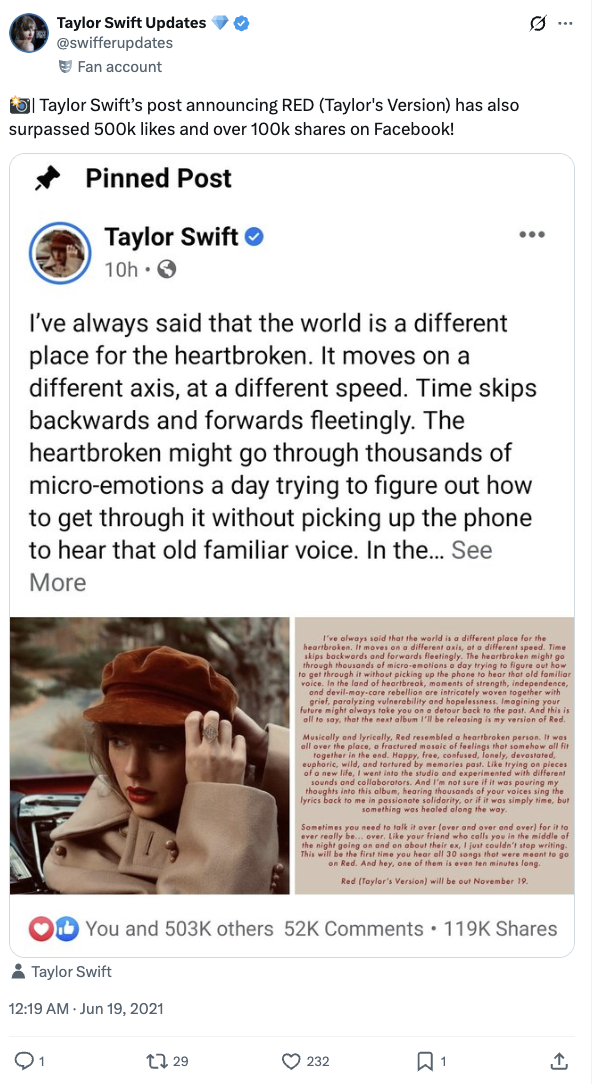}
    \Description{...}
    \caption{A Twitter account (@swifferupdates) \href{https://x.com/swifferupdates/status/1406104309930938370}{shared} a screenshot of Taylor Swift's Facebook post on their Twitter page.}
    \label{fbONtweet1}
\end{figure}

\begin{figure}[H]
    \includegraphics[scale=0.43]{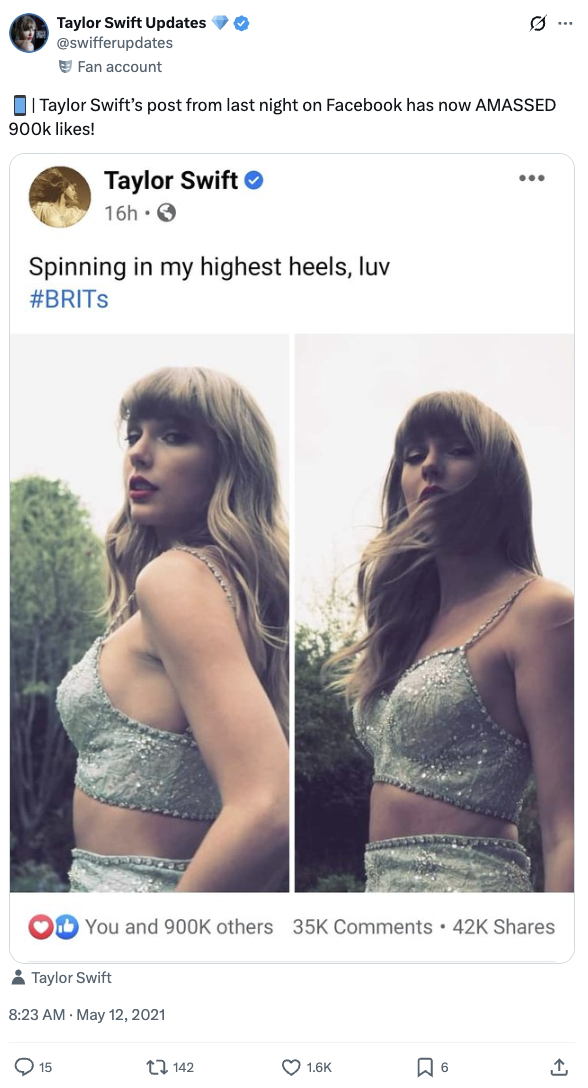}
    \Description{...}
    \caption{A Twitter account (@swifferupdates) \href{https://x.com/swifferupdates/status/1392455481860317185}{shared} a screenshot of Taylor Swift's Facebook post on their Twitter page.}
    \label{fbONtweet2}
\end{figure}

Figure \ref{fbONtweet3} shows an example where a Twitter account \href{https://x.com/TSwiftVS/status/1372024149145649154}{@TSwiftVS} shared a screenshot of \href{https://www.facebook.com/TaylorSwift/}{Taylor Swift's} Facebook post on their Twitter page. Figures \ref{fbONtweet1} and \ref{fbONtweet2} show examples where a Twitter account \href{https://x.com/swifferupdates}{@swifferupdates} shared screenshots of \href{https://www.facebook.com/TaylorSwift/}{Taylor Swift's} Facebook post on their Twitter page.

\FloatBarrier
\clearpage

\section{Screenshots of Facebook posts shared on Instagram}
\label{app:fbONinsta}
\begin{figure}[H]
    \includegraphics[scale=0.35]{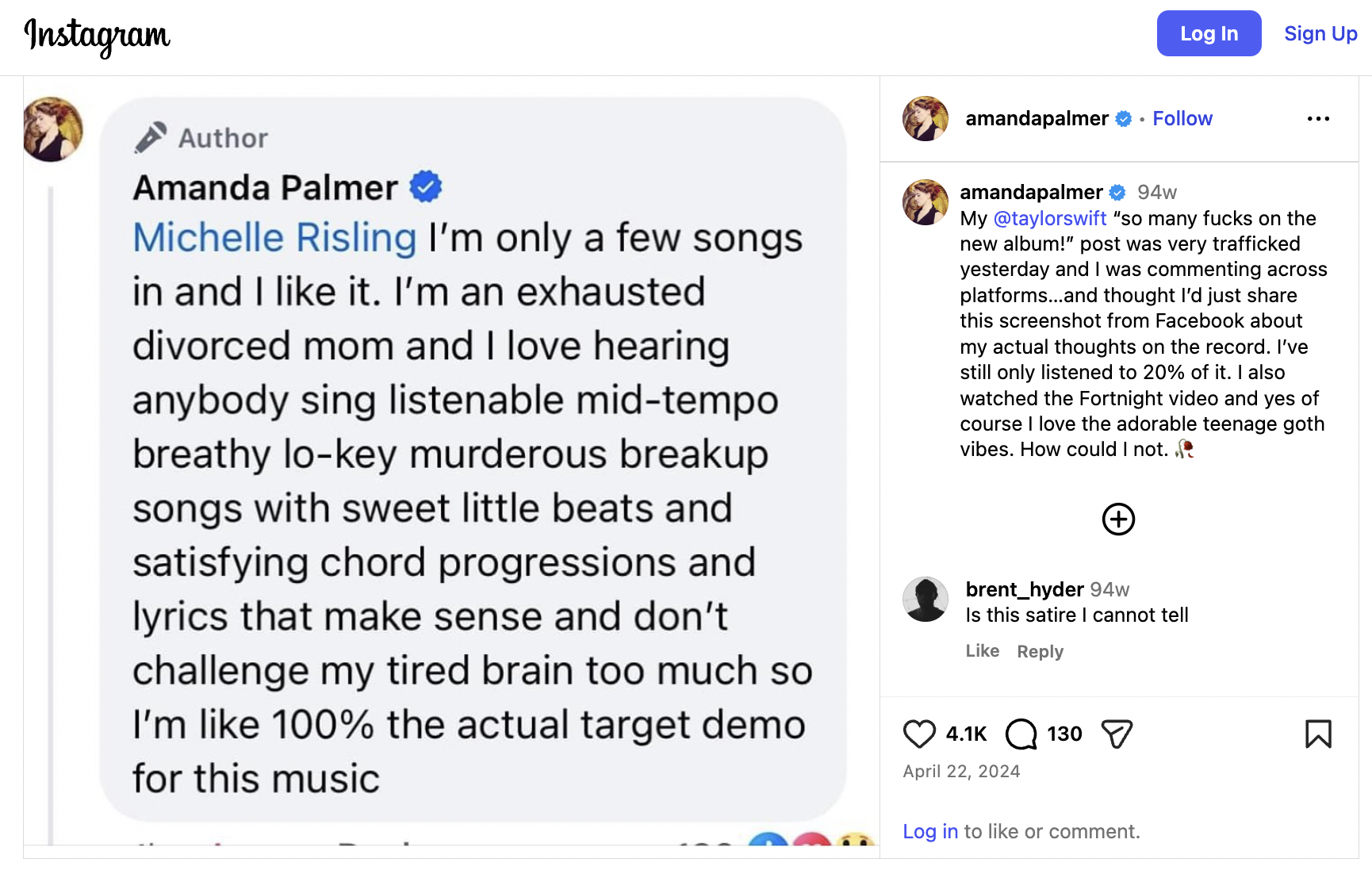}
    \Description{...}
    \caption{An Instagram account (amandapalmer) \href{https://www.instagram.com/p/C6EJDwkLxs9/}{shared} a screenshot of her own Facebook post on her Instagram page.}
    \label{fbONinsta3}
\end{figure}

\begin{figure}[H]
    \includegraphics[scale=0.4]{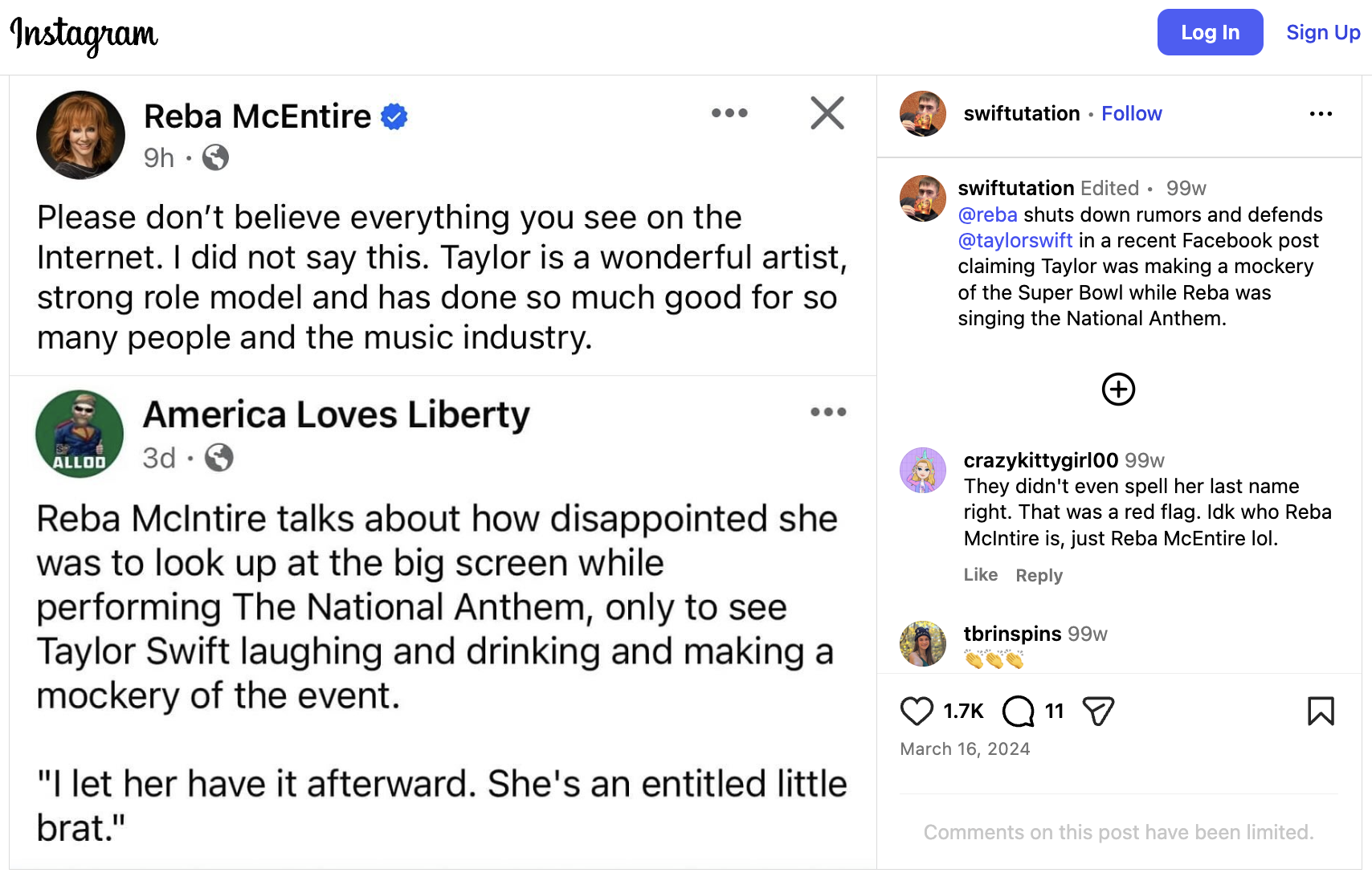}
    \Description{...}
    \caption{An Instagram account (swiftutation) \href{https://www.instagram.com/p/C4mAYwcuMTq/}{shared} a screenshot of Reba McEntire's Facebook post on their Instagram page.}
    \label{fbONinsta1}
\end{figure}

Figure \ref{fbONinsta3} shows an example where an Instagram account \href{https://www.instagram.com/amandapalmer}{amandapalmer} shared a screenshot of her own Facebook post on her Instagram page. Figures \ref{fbONinsta1} and \ref{fbONinsta3} show examples where an Instagram account \href{https://www.instagram.com/swiftutation}{swiftutation} shared a screenshot of \href{https://www.facebook.com/reba/}{Reba McEntire} and \href{https://www.facebook.com/sabrinacarpenter/}{Sabrina Carpenter's} Facebook post on their Instagram page.

\begin{figure}[H]
    \includegraphics[scale=0.35]{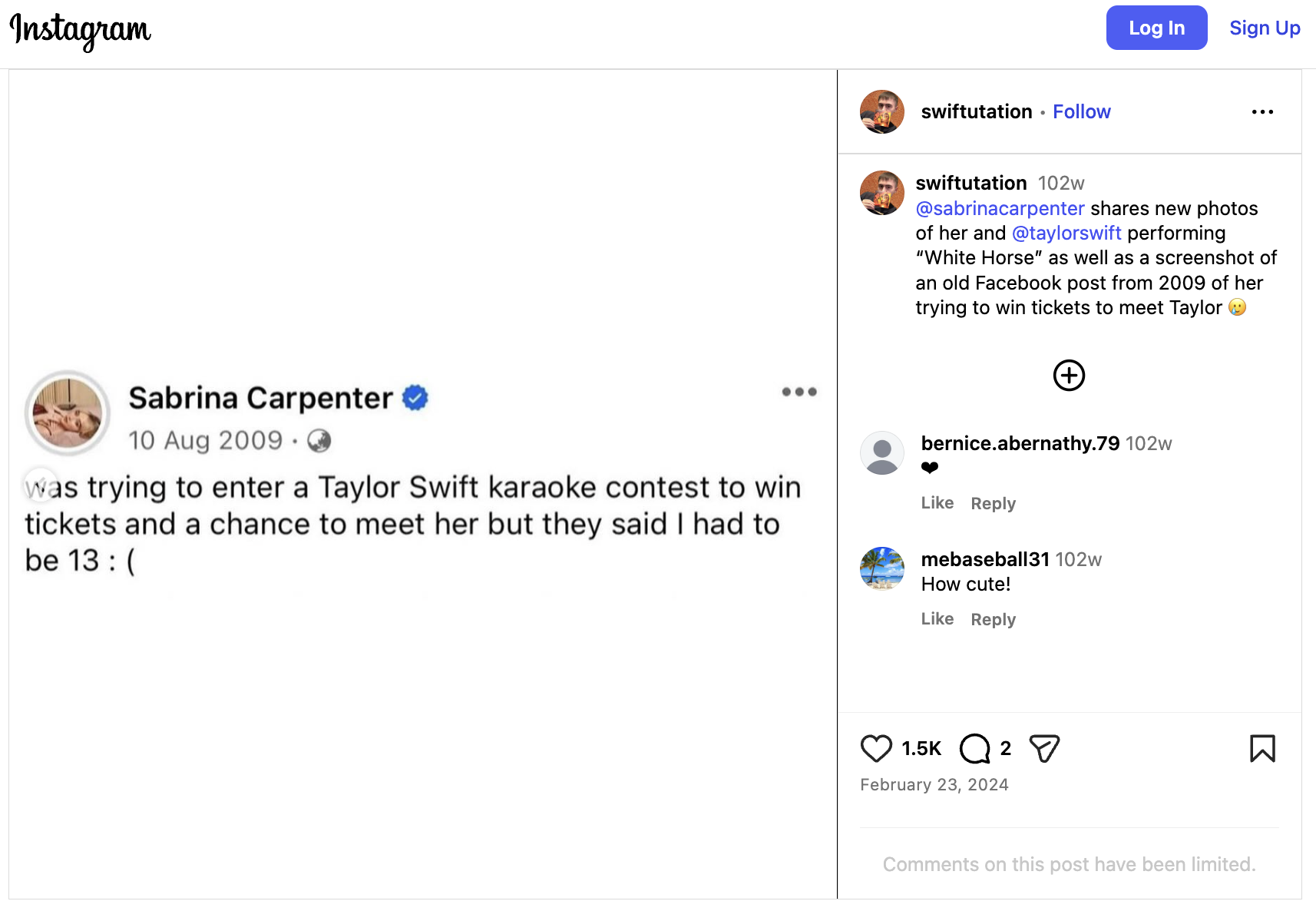}
    \Description{...}
    \caption{An Instagram account (swiftutation) \href{https://www.instagram.com/p/C3tpV2hO8-C/?img_index=3}{shared} a screenshot of Sabrina Carpenter's Facebook post on their Instagram page.}
    \label{fbONinsta2}
\end{figure}

\FloatBarrier
\clearpage

\section{Screenshots of an Instagram post shared on Twitter}
\label{app:instaONtweet}
\begin{figure}[H]
    \includegraphics[scale=0.8]{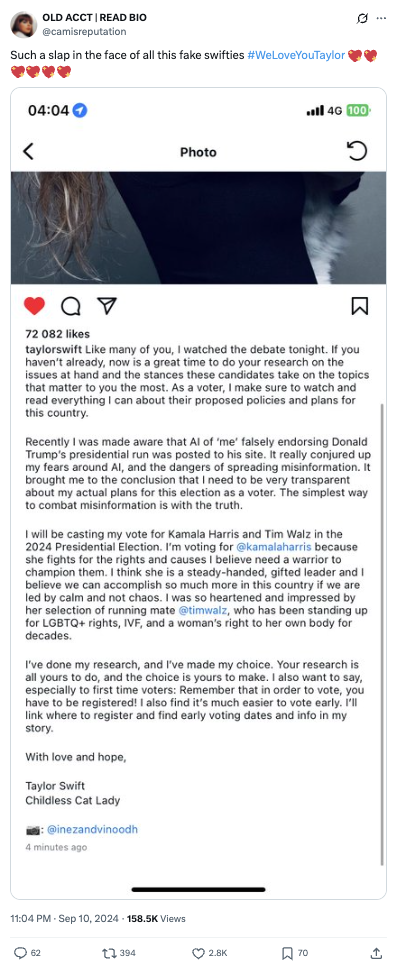}
    \Description{...}
    \caption{A Twitter account (camisreputation) \href{https://x.com/camisreputation/status/1833703209677869310}{shared} a screenshot of Taylor Swift's Instagram post on their Twitter page.}
    \label{instaONtweet1}
\end{figure}

\begin{figure}[H]
    \includegraphics[scale=0.4]{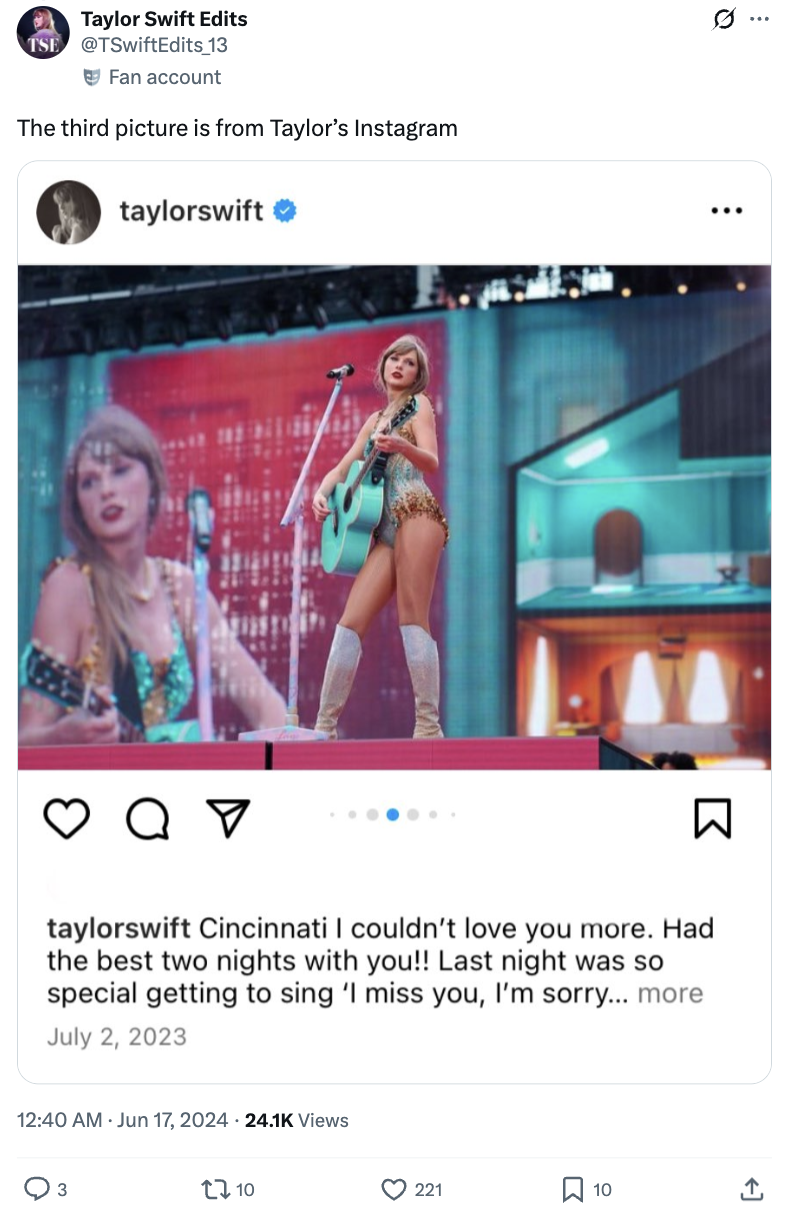}
    \Description{...}
    \caption{A Twitter account (@TSwiftEdits\_13) \href{https://x.com/TSwiftEdits\_13/status/1802561961898680746}{shared} a screenshot of Taylor Swift's Instagram post on their Twitter page.}
    \label{instaONtweet2}
\end{figure}

Figure \ref{instaONtweet1} shows an example where a Twitter account \href{https://x.com/camisreputation}{camisreputation} shared a screenshot of \href{https://www.instagram.com/taylorswift/}{Taylor Swift's} Instagram post on their Twitter page. Figure \ref{instaONtweet2} shows another example where a Twitter account \href{https://x.com/TSwiftEdits\_13}{@TSwiftEdits\_13} shared a screenshot of \href{https://www.instagram.com/taylorswift/}{Taylor Swift's} Instagram post on their Twitter page. Another example in Figure \ref{instaONtweet1} shows a Twitter account \href{https://x.com/selovelenaa}{@selovelenaa} shared a screenshot of \href{https://www.instagram.com/taylorswift/}{Taylor Swift's} Instagram post on their Twitter page.

\begin{figure}[H]
    \includegraphics[scale=0.5]{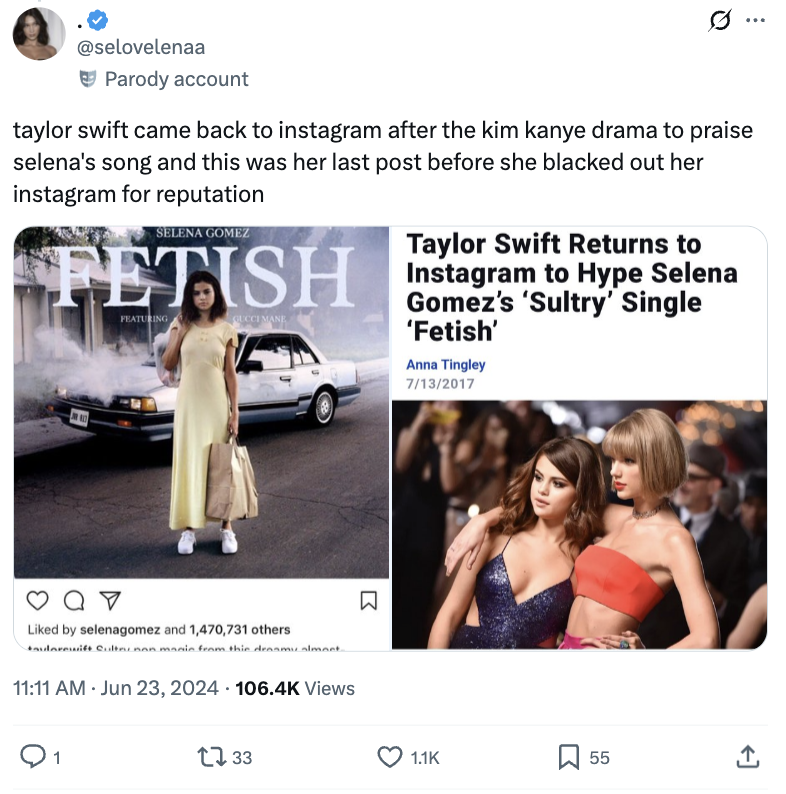}
    \Description{...}
    \caption{An Twitter account (@selovelenaa) \href{https://x.com/selovelenaa/status/1804895168203751866}{shared} a screenshot of Taylor Swift's Instagram post on their Twitter page.}
    \label{instaONtweet3}
\end{figure}

\FloatBarrier
\clearpage

\section{Screenshots of an Instagram post shared on Facebook}
\label{app:instaONfb}
\begin{figure}[H]
    \includegraphics[scale=0.4]{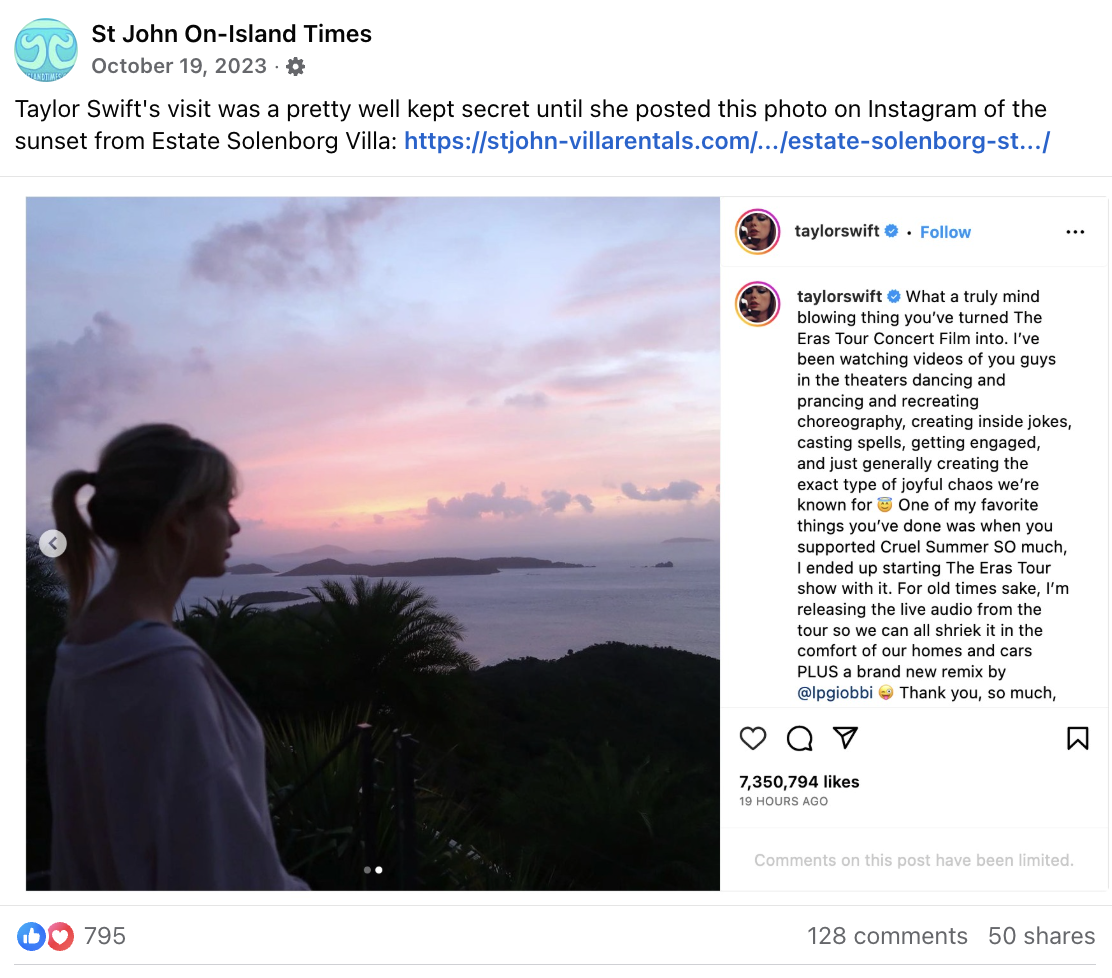}
    \Description{...}
    \caption{A Facebook account (St John On-Island Times) \href{https://www.facebook.com/permalink.php?story_fbid=pfbid0T5qeYeweNkuHdpPUWh5bUFHCfY28y8wRLWYcBqj193eVmGnZYQJmMfLwPR2535Mdl&id=100064823915472}{shared} a screenshot of Taylor Swift's Instagram post on their Facebook page.}
    \label{instaONfb1}
\end{figure}

\begin{figure}[H]
    \includegraphics[scale=0.45]{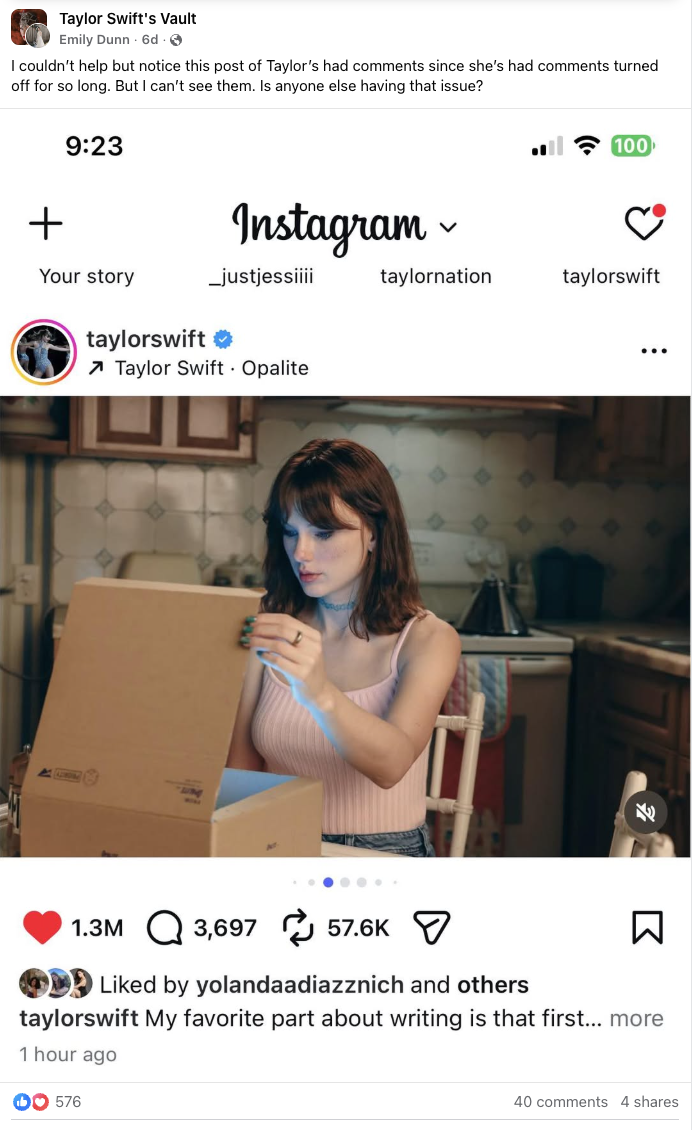}
    \Description{...}
    \caption{A Facebook account (Taylor Swift's Vault) \href{https://www.facebook.com/groups/2254218764714763/posts/3958480274288595/}{shared} a screenshot of Taylor Swift's Instagram post on their Facebook page.}
    \label{instaONfb2}
\end{figure}

\begin{figure}[H]
    \includegraphics[scale=0.45]{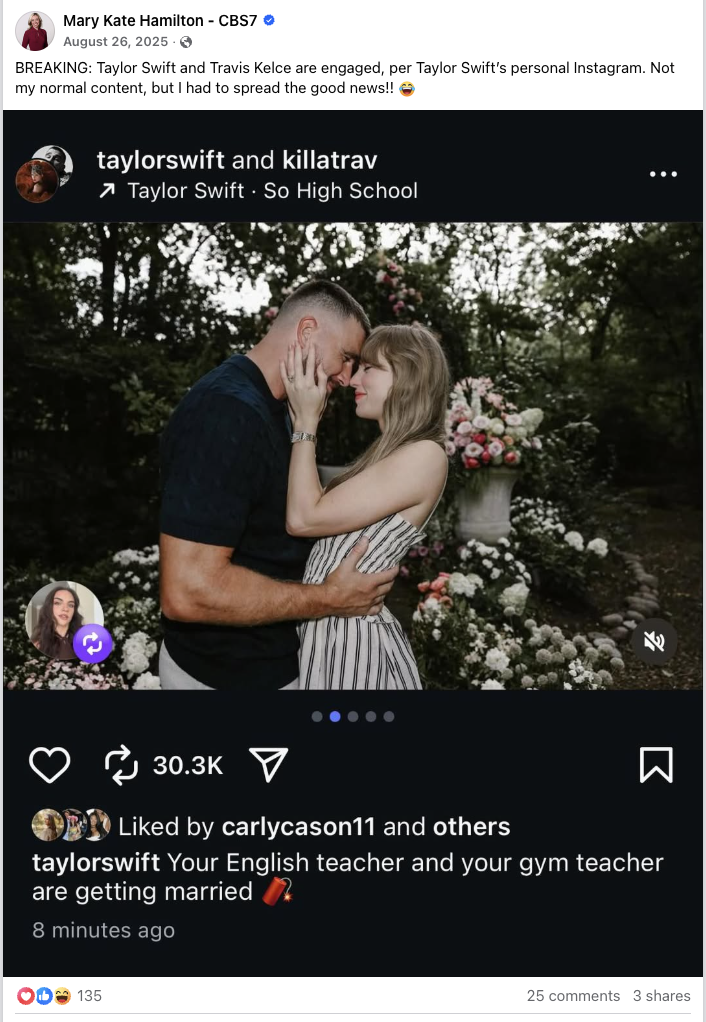}
    \Description{...}
    \caption{An Facebook account (Mary Kate Hamilton - CBS7) \href{https://www.facebook.com/MaryKateHamiltonCBS7/posts/pfbid02LK7rWe1vXgGGsiNp4XvVavtVV9pC93hcgJwKNkiwNSsgqtvDtg8dfKPvaRoe3tWNl}{shared} a screenshot of Taylor Swift's Instagram post on their Facebook page.}
    \label{instaONfb3}
\end{figure}

Figure \ref{instaONfb1} shows an example where a Facebook account \href{https://www.facebook.com/people/St-John-On-Island-Times}{St John On-Island Times} shared a screenshot of \href{https://www.instagram.com/taylorswift/}{Taylor Swift's} Instagram post on their Facebook page. Figure \ref{instaONfb2} shows another example where a Facebook account \href{https://www.facebook.com/groups/2254218764714763/}{Taylor Swift's Vault} shared a screenshot of \href{https://www.instagram.com/taylorswift/}{Taylor Swift's} Instagram post on their Facebook page. Figure \ref{instaONfb3} shows another example where a Facebook account \href{https://www.facebook.com/MaryKateHamiltonCBS7}{Mary Kate Hamilton - CBS7} shared a screenshot of \href{https://www.instagram.com/taylorswift/}{Taylor Swift's} Instagram post on their Facebook page.

\FloatBarrier
\clearpage

\section{Aggregating screenshots}
\label{app:aggregate_ss}
\begin{figure}[H]
    \includegraphics[scale=0.45]{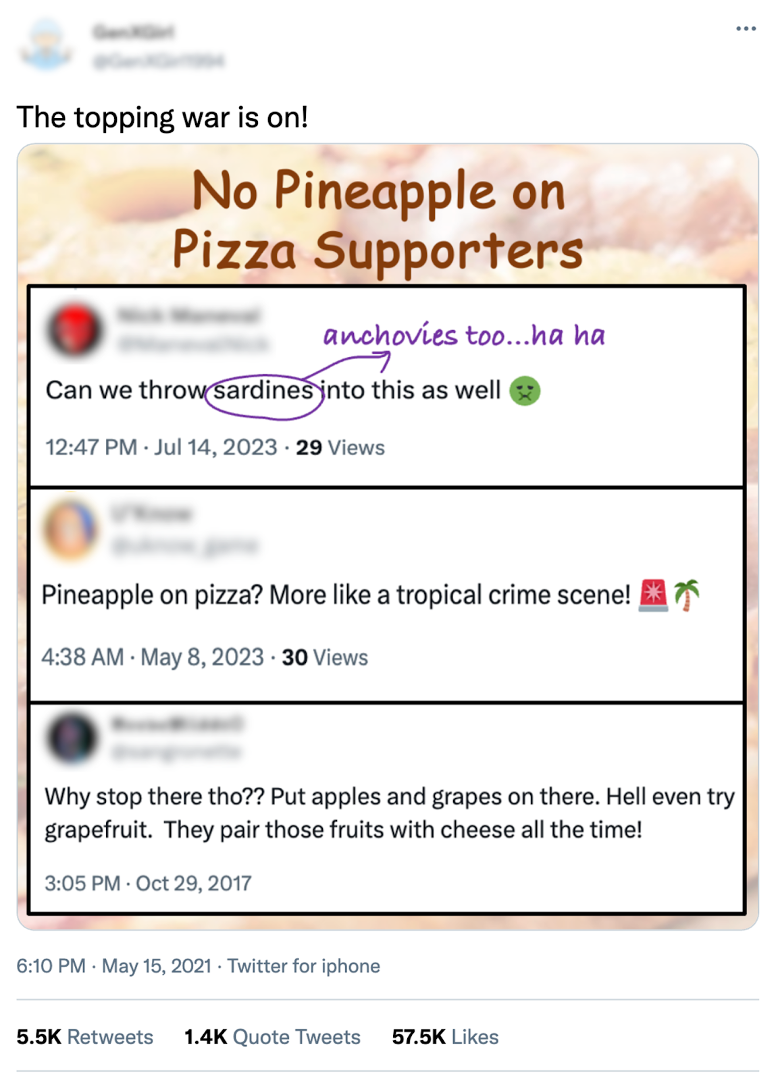}
    \Description{...}
    \caption{An aggregated screenshot showing three anti-pineapple tweets stitched together.}
    \label{aggregate_ss}
\end{figure}

\begin{figure}[H]
    \includegraphics[scale=0.45]{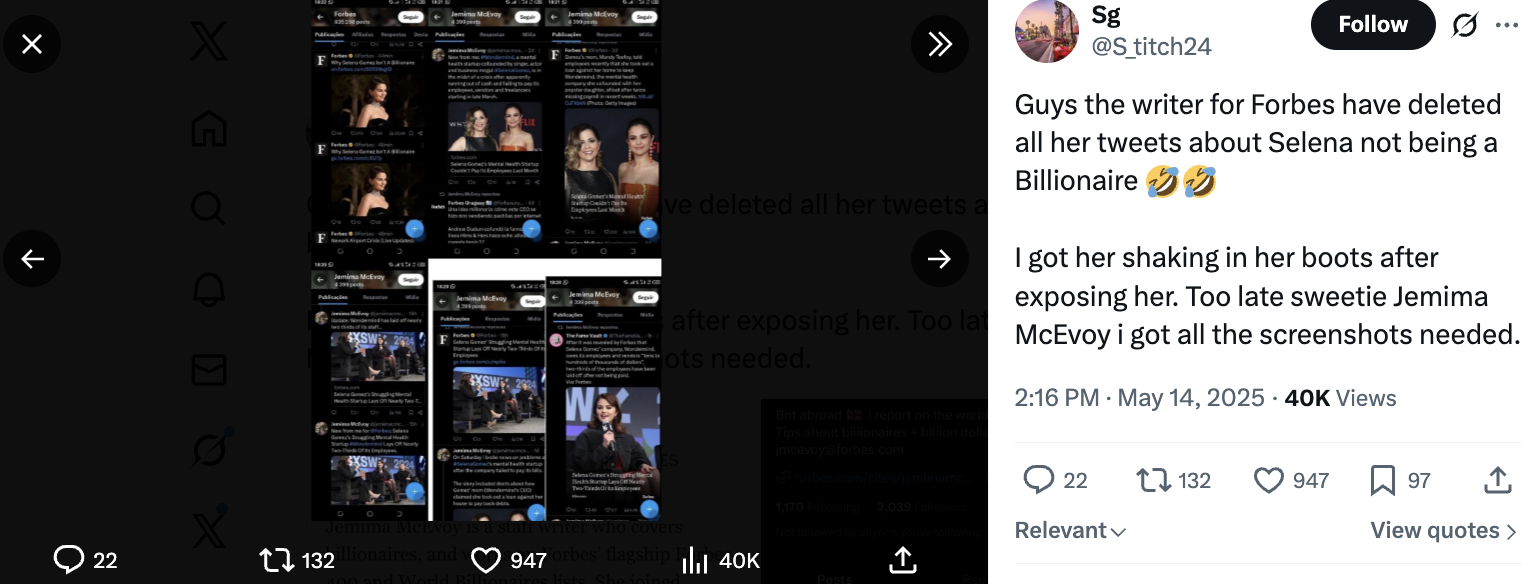}
    \Description{...}
    \caption{An aggregated screenshot showing multiple screenshots of tweets related to Selena Gomez.}
    \label{selena}
\end{figure}

\begin{figure}[H]
    \includegraphics[scale=0.45]{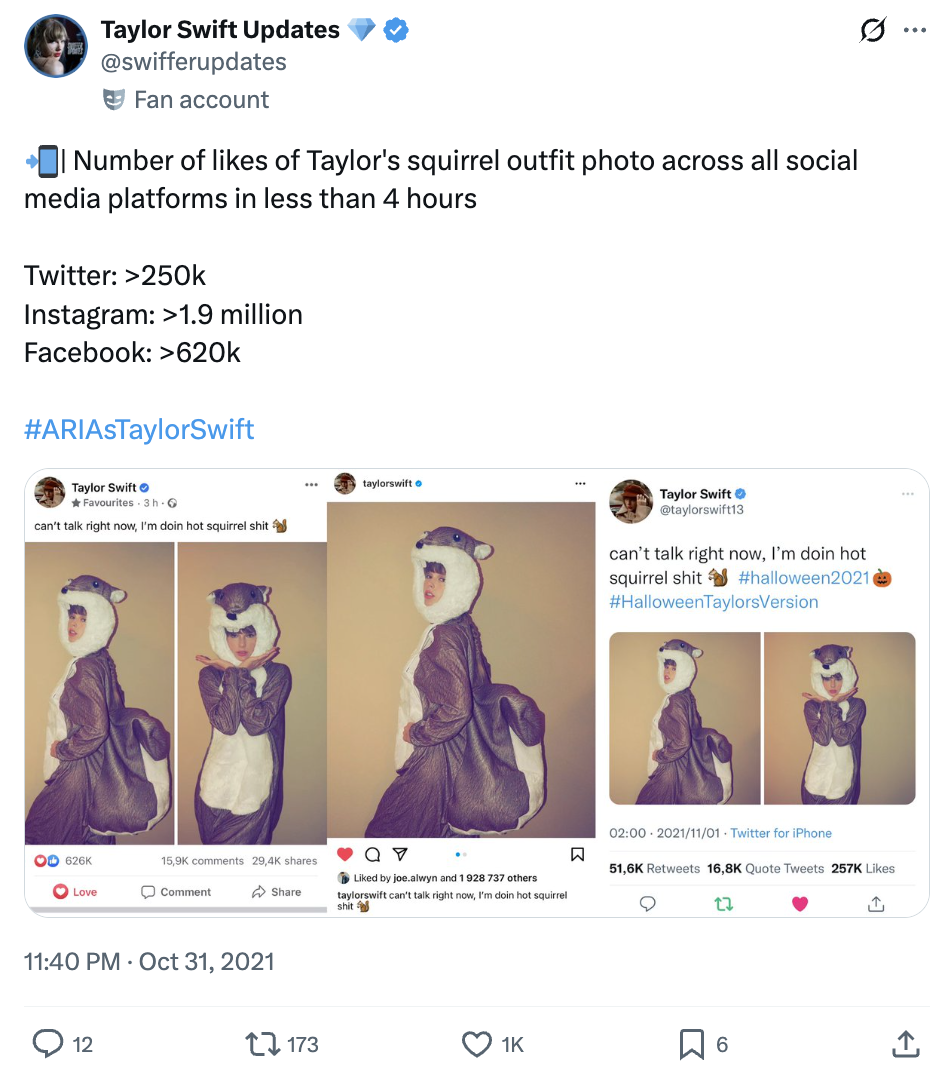}
    \Description{...}
    \caption{An aggregated screenshot showing same post of Taylor Swift on Twitter, Instagram, and Facebook shared on Twitter.}
    \label{taylor}
\end{figure}

This section provides additional screenshot examples for cross-platform sharing. Figure \ref{aggregate_ss} shows three tweets stitched together and shared as a screenshot to present a narrative supporting no pineapple on pizza. Figure \ref{selena} shows a screenshot \href{https://x.com/S_titch24/status/1922717642546143504/photo/3}{shared on Twitter} consisting of multiple screenshot of tweets stitched together related to Selena Gomez. Figure \ref{taylor} shows an example of screenshot \href{https://x.com/swifferupdates/status/1455016906356838403}{shared on Twitter} consisting of Taylor Swift's same post on Twitter, Instagram, and Facebook stitched together.

\FloatBarrier
\clearpage

\section{Satire and humor}
\label{app:satire/humor}
\begin{figure}[H]
    \includegraphics[scale=0.4]{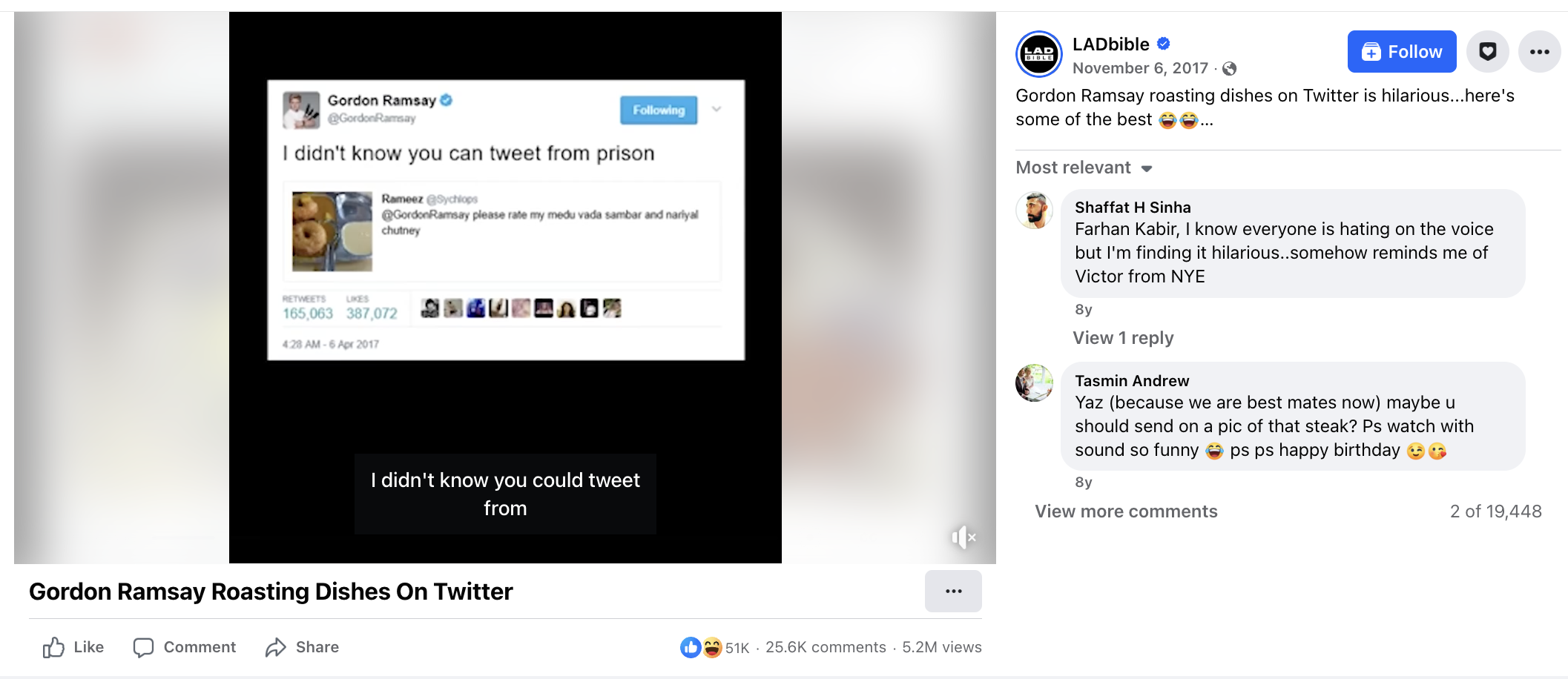}
    \Description{...}
    \caption{A \href{https://www.facebook.com/LADbible/videos/3992982604082233/}{video} that combines screenshots of Gordon Ramsay's roasting and savage tweet replies by a Facebook account \href{https://www.facebook.com/LADbible}{LADbible}.}
    \label{satire2}
\end{figure}

\begin{figure}[H]
    \includegraphics[scale=0.4]{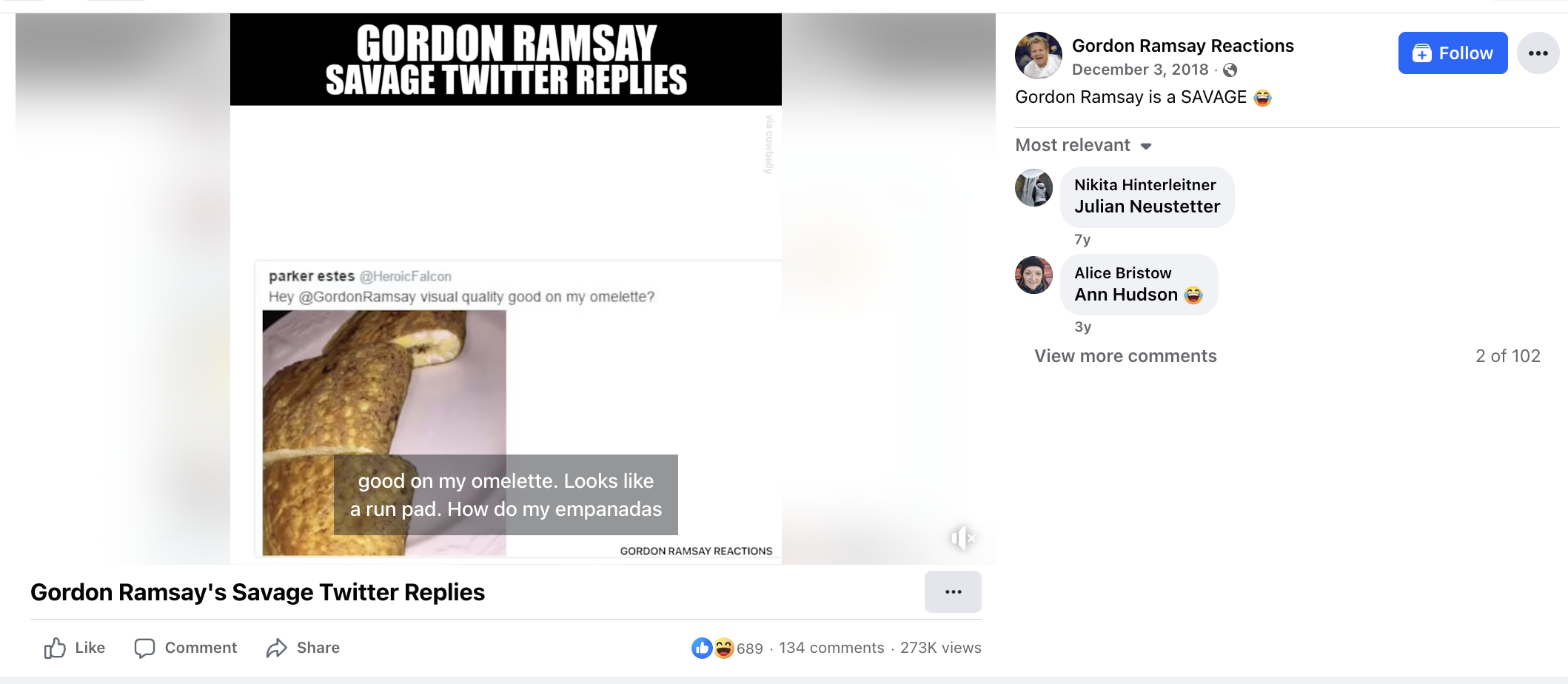}
    \Description{...}
    \caption{A \href{https://www.facebook.com/GordonRamsayReactions/videos/gordon-ramsays-savage-twitter-replies/358945088197183/}{video} that combines screenshots of Gordon Ramsay's roasting and savage tweet replies by the Facebook account \href{https://www.facebook.com/GordonRamsayReactions}{Gordon Ramsay Reactions}.}
    \label{satire3}
\end{figure}

This section provides additional screenshot examples for satire/humor purpose. Figures \ref{satire2} and \ref{satire3} show a video consisting of screenshots of Gordon Ramsay's roasting and savage tweet replies are used for satire/humor purpose.

\FloatBarrier
\clearpage

\section{Enabling commentary and annotation}
\label{app:annotate}
\begin{figure}[H]
    \includegraphics[scale=0.45]{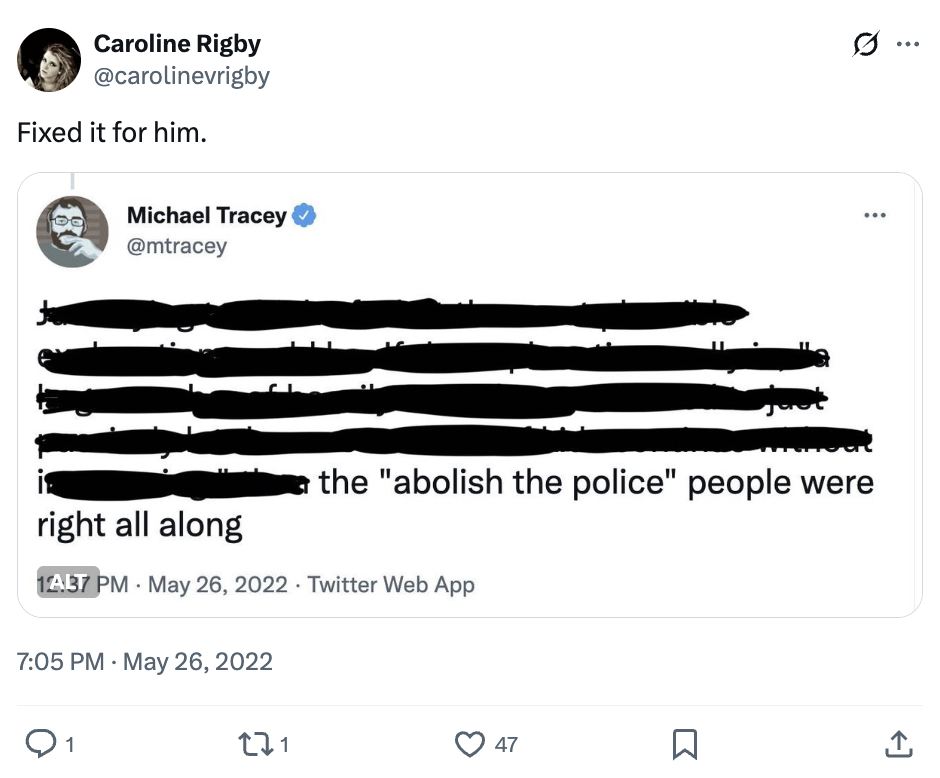}
    \Description{...}
    \caption{Screenshot of a tweet \href{https://x.com/carolinevrigby/status/1529961960044912641}{shared} on Twitter, annotated with black indicating redaction.}
    \label{black}
\end{figure}

\begin{figure}[H]
    \includegraphics[scale=0.45]{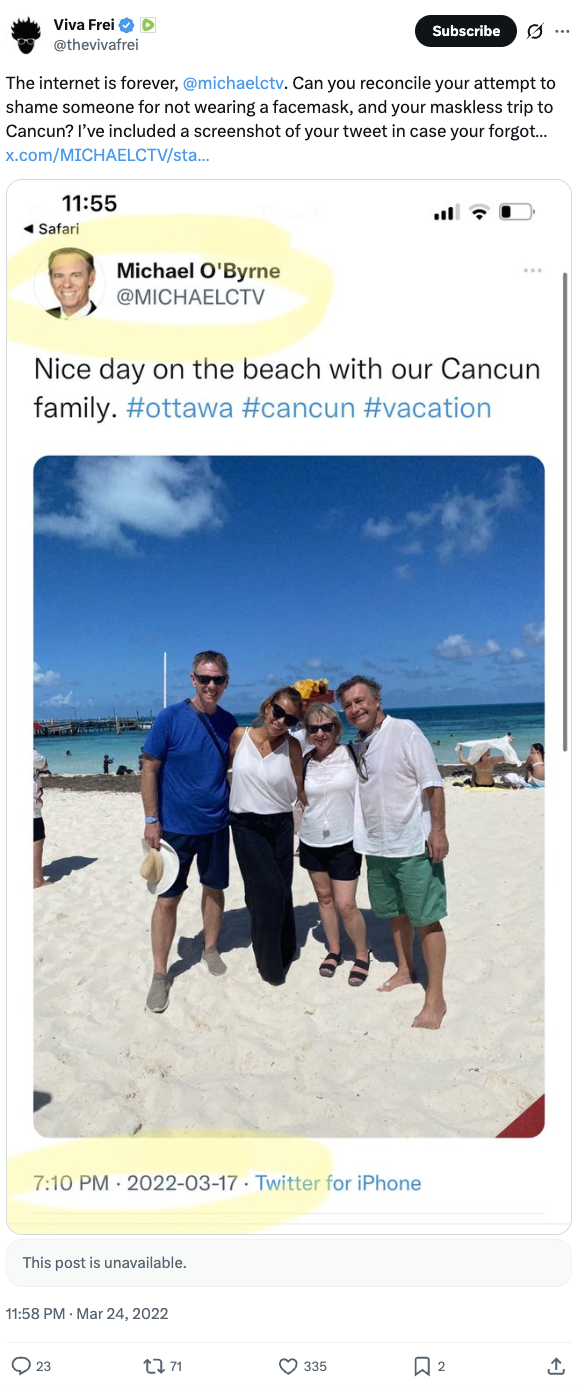}
    \Description{...}
    \caption{Screenshot of a tweet \href{https://x.com/thevivafrei/status/1507205276704944136}{shared} on Twitter, annotated with yellow highlighted colored circles.}
    \label{yellow}
\end{figure}

This section provides additional screenshot examples for enabling commentary and annotation. Figure \ref{black} shows screenshot of a tweet shared on Twitter, annotated with black indicating redaction. Figure \ref{yellow} shows screenshot of a tweet shared on Twitter, annotated with yellow highlighted colored circles.

\FloatBarrier
\clearpage

\section{Denying engagement}
\label{app:deny_engage}
\begin{figure}[H]
    \includegraphics[scale=0.45]{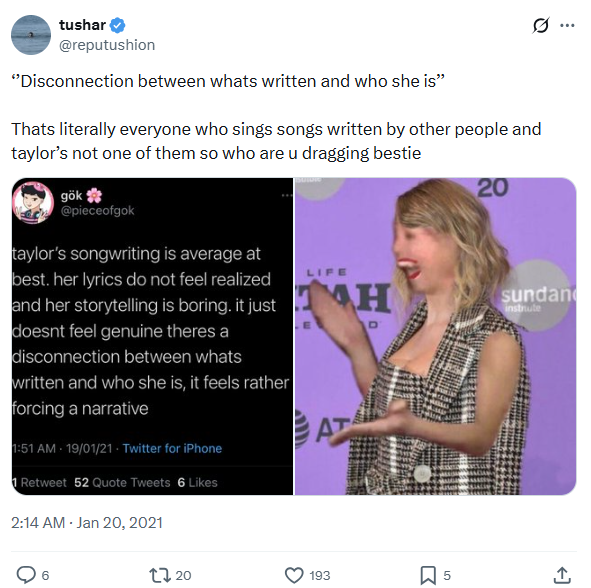}
    \Description{...}
    \caption{ By \href{https://x.com/reputushion/status/1351790067002482689}{sharing} a screenshot of @pieceofgok's tweet, @reputushion criticized a statement about Taylor Swift without directly engaging with the original post.}
    \label{deny_engage1}
\end{figure}

\begin{figure}[H]
    \includegraphics[scale=0.45]{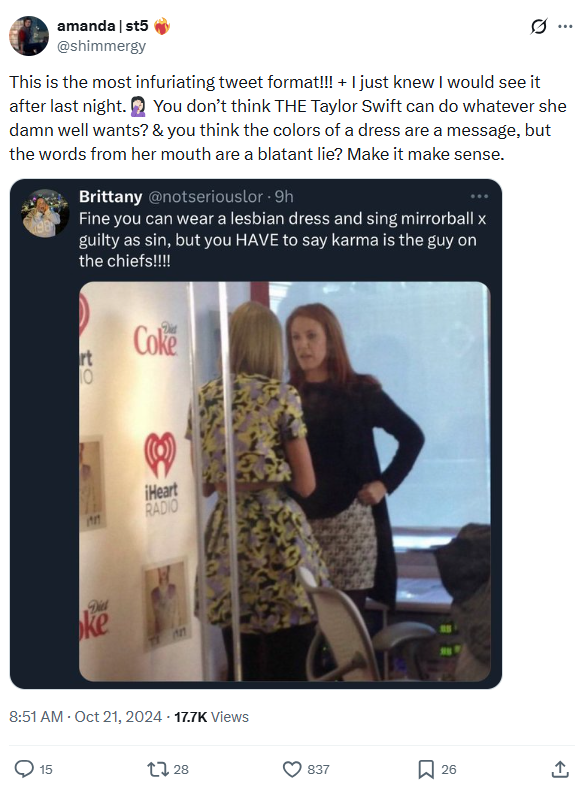}
    \Description{...}
    \caption{By \href{https://x.com/shimmergy/status/1848346416772534473}{sharing} a screenshot of @notseriouslor tweet, @shimmergy criticized a statement about Taylor Swift without directly engaging with the original post.}
    \label{deny_engage2}
\end{figure}

This section provides additional screenshot examples for denying engagement. Figure \ref{deny_engage1} shows a Twitter user \href{https://x.com/reputushion}{@reputushion} shared a screenshot of another user \href{https://x.com/pieceofgok}{@pieceofgok's} tweet criticizing Taylor Swift. Another similar example is shown in Figure \ref{deny_engage2} where a Twitter user \href{https://x.com/shimmergy}{@shimmergy} shared a screenshot of another user \href{https://x.com/notseriouslor}{@notseriouslor's} tweet criticizing Taylor Swift. Both examples show how users indirectly engage with a post by denying platform-specific engagement features.

\end{document}